\pdfoutput=1
\documentclass[a4paper,UKenglish,cleveref, autoref]{lipics-v2019}

\usepackage{amsmath}
\usepackage{MnSymbol}
\usepackage{algorithm}
\usepackage[noend]{algpseudocode}
\usepackage{varwidth}
\usepackage{thmtools}
\usepackage{todonotes}
\usepackage{multicol}
\usepackage{dsfont}

\usepackage{caption}
\usepackage{subcaption}

\usepackage{lipsum}
\usepackage{graphicx}

\usepackage{tikz}
\usepackage{pgfplots}

\pgfplotsset{compat=1.12}
\usetikzlibrary{arrows,decorations.pathmorphing,backgrounds,positioning,fit,calc,automata}
\usetikzlibrary{trees}
\usetikzlibrary{shapes}
\usetikzlibrary{chains}
\usetikzlibrary{patterns}

\newcommand{\delay}{\operatorname{delay}}

\newcommand{\dom}{\operatorname{dom}}
\newcommand{\supp}{\operatorname{supp}}
\newcommand{\enc}{\operatorname{enc}}

\newcommand{\costf}{\kappa}
\newcommand{\costt}[1]{\operatorname{cost}_{#1}}
\newcommand{\RENUM}{\textsc{RANK-ENUM}}
\newcommand{\RENUMT}{\textsc{RANK-ENUM-T}}
\newcommand{\RAENUM}{\textsc{RA-ENUM}}
\newcommand{\RENUMCEA}{\textsc{RANK-ENUM-CEA}}

\newcommand{\X}{\bar{X}}
\newcommand{\x}{\bar{x}}

\newcommand{\bbG}{\mathbb{G}}
\newcommand{\gop}{\oplus}
\newcommand{\biggop}{\bigoplus}
\newcommand{\gid}{\mathds{O}}
\newcommand{\gsum}{\Sigma\,}

\newcommand{\sop}{\odot}

\newcommand{\ksum}{\oplus}
\newcommand{\kprod}{\odot}
\newcommand{\bigksum}{\bigoplus}
\newcommand{\bigkprod}{\bigodot}
\newcommand{\kzero}{\mathtt{0}}
\newcommand{\kone}{\mathtt{1}}

\newcommand{\bbN}{\mathbb{N}}
\newcommand{\bbZ}{\mathbb{Z}}
\newcommand{\cA}{\mathcal{A}}

\newcommand{\cL}{\mathcal{L}}

\newcommand{\cO}{\mathcal{O}}

\newcommand{\cT}{\mathcal{T}}

\newcommand{\sem}[1]{{\lsem{}{#1}\rsem}}
\newcommand{\trans}[2][]{\raisebox{-1pt}[10pt][0pt]{$\overset{#2}{\underset{^{#1}}{\raisebox{0pt}[3pt][0pt]{$\relbar\mspace{-8mu}\longrightarrow$}}}$}}
\newcommand{\amark}{\bullet}
\newcommand{\umark}{\circ}

\newcommand{\run}{\operatorname{Run}}

\newcommand{\out}{\operatorname{out}}

\newcommand{\Open}[1]{#1~\mkern-12mu\vdash}
\newcommand{\Close}[1]{\dashv~\mkern-10mu#1}
\newcommand{\mspan}[1]{[#1\rangle}
\newcommand{\xset}{\textbf{X}}

\newcommand{\dset}{\mathbb{D}}
\newcommand{\aset}{\mathbb{A}}

\newcommand{\cR}{\mathcal{R}}
\newcommand{\att}{\operatorname{att}}
\newcommand{\type}{\operatorname{type}}
\newcommand{\tuples}{\operatorname{tuples}}
\newcommand{\upset}{\mathcal{U}}

\newcommand{\WITHIN}{\ \texttt{WITHIN} \ }

\newcommand{\madd}{{\tt add}}

\renewcommand{\path}{\operatorname{path}}
\newcommand{\mfindmin}{{\sf findMin}}
\newcommand{\misempty}{{\sf isEmpty}}
\newcommand{\mdeletemin}{{\sf deleteMin}}
\newcommand{\mmeld}{{\sf meld}}
\newcommand{\mincreaseby}{{\sf increaseBy}}
\newcommand{\mminPriority}{{\sf minPrio}}

\newcommand{\iisempty}{\textsf{isEmpty}}

\newcommand{\mlink}{{\tt link}}

\algnewcommand\algorithmicforeach{\textbf{for each}}
\algdef{S}[FOR]{ForEach}[1]{\algorithmicforeach\ #1\ \algorithmicdo}

\algrenewcommand\algorithmicrequire{\textbf{input:}}
\algrenewcommand\algorithmicensure{\textbf{output:}}

\newcommand{\hname}{heap of words}
\newcommand{\hnameshort}{HoW}
\newcommand{\hnameshortp}{HoWs}
\newcommand{\hnamebigcaps}{Heap of Words}
\newcommand{\hstruct}{\hnode}

\newcommand{\hnode}{h}
\newcommand{\hpair}[2]{[#1\!:\!#2]}

\newcommand{\hempty}{\emptyset}
\newcommand{\hfindmin}{\textsc{FindMin}}

\newcommand{\hdeletemin}{\textsc{DeleteMin}}
\newcommand{\hmeld}{\textsc{Meld}}
\newcommand{\hadd}{\textsc{Add}}
\newcommand{\hincreaseby}{\textsc{IncreaseBy}}
\newcommand{\hextendby}{\textsc{ExtendBy}}

\newcommand{\Run}{\operatorname{Run}}

\newcommand{\sDAG}{{string-DAG}}
\newcommand{\fprio}{\operatorname{prioritize}}
\newcommand{\hq}[1]{\langle#1\rangle}

\newcommand{\qpair}[2]{[#1\!:\!#2]}

\newcommand{\set}[1]{[#1]}

\newcommand{\tT}{\texttt{T}}

\newcommand{\tdom}{V}
\newcommand{\tdomi}[1]{\tdom_{#1}}
\newcommand{\fch}{\textsf{first}}
\newcommand{\fchi}[1]{\fch_{#1}}
\newcommand{\nsb}{\textsf{next}}
\newcommand{\nsbi}[1]{\nsb_{#1}}
\newcommand{\troot}{v^0}
\newcommand{\trooti}[1]{\troot_{#1}}
\newcommand{\troots}{\textsf{roots}}
\newcommand{\trootsi}[1]{\troots_{#1}}
\newcommand{\chd}{\textsf{children}}
\newcommand{\chdi}[1]{\chd_{#1}}
\newcommand{\parent}{\textsf{parent}}

\newcommand{\op}{\textsf{op}}
\newcommand{\opi}[1]{\op_{#1}}

\newcommand{\tempty}{\bot}
\newcommand{\trank}{\text{rank}}
\newcommand{\tdelta}{\Delta}
\newcommand{\tdeltai}[1]{\tdelta_{#1}}
\newcommand{\tprio}{\textsf{pr}}
\newcommand{\tprioi}[1]{\tprio_{#1}}
\newcommand{\tval}{\textsf{elem}}
\newcommand{\tvali}[1]{\tval_{#1}}
\newcommand{\tdeltainit}{\delta^0}
\newcommand{\tdeltainiti}[1]{\tdeltainit_{#1}}

\newcommand{\M}{M}

\newcommand{\T}{T}
\newcommand{\h}{H}

\title{Ranked enumeration of MSO logic on words} 

\titlerunning{Ranked enumeration of MSO logic on words}

\author{Pierre Bourhis}{CNRS Lille, CRIStAL UMR 9189, University of Lille, INRIA Lille, France}{}{}{}

\author{Alejandro Grez}{PUC \& IMFD, Chile}{}{}{}

\author{Louis Jachiet}{LTCI, IP Paris, France}{}{}{}

\author{Cristian Riveros}{PUC \& IMFD, Chile}{}{}{}

\authorrunning{P. Bourhis, A. Grez, L. Jachiet, and C. Riveros}

\Copyright{Pierre Bourhis, Alejandro Grez, Louis Jachiet, and Cristian Riveros}



\nolinenumbers 

\hideLIPIcs  

\begin{document}
	
\maketitle

\begin{abstract}

In the last years, enumeration algorithms with bounded delay have attracted a lot of attention for several data management tasks. Given a query and the data, the task is to preprocess the data and then enumerate all the answers to the query one by one and without repetitions. This enumeration scheme is typically useful when the solutions are treated on the fly or when we want to stop the enumeration once the pertinent solutions have been found. However, with the current schemes, there is no restriction on the order how the solutions are given and this order usually depends on the techniques used and not on the relevance for the user.

In this paper we study the enumeration of monadic second order logic (MSO) over words when the solutions are ranked. We present a framework based on \emph{MSO cost functions} that allows to express MSO formulae on words with a cost associated with each solution. We then demonstrate the generality of our framework which subsumes, for instance, document spanners and regular complex event processing queries and adds ranking to them. The main technical result of the paper is an algorithm for enumerating all the solutions of formulae in increasing order of cost efficiently, namely, with a linear preprocessing phase and logarithmic delay between solutions. The novelty of this algorithm is based on using functional data structures, in particular, by extending functional Brodal queues to suit with the ranked enumeration of MSO on words.
\end{abstract}

\section{Introduction} \label{sec:introduction}

Managing and querying word structures such as texts has been one of the classical problems of different communities in computer science. In particular, this problem has been predominant in information extraction where the goal is to extract some subparts of a text. A logical approach that has brought a lot of attention in the database community is document spanners~\cite{FaginKRV15}. This logical framework provides a language for extracting subparts of a document. More specifically, regular spanners are based on regular expressions that fill relations with tuples of the texts' subparts. These relations can afterwards be queried by conjunctive or datalog-like queries. 

The document spanners' main algorithmic problem is the efficient evaluation of a spanner over a word. Recently, a novel approach has been to focus on the enumeration problem to obtain efficient evaluation algorithms. The principle of an enumeration algorithm is to create a representation of the set of answers efficiently depending only on the input word's size and the query, and not in the number of answers. This time is called \emph{the preprocessing time}. The second part of an enumeration algorithm is to enumerate the outputs one by one using the previous representation. The time between two consecutive outputs is called the \emph{delay}. As for the preprocessing time, an efficient delay should not depend on the number of outputs, but only on the input size (i.e., word and query). In general, the most efficient enumeration algorithms have linear preprocessing time and constant delay, both in the size of the input.

Several people have studied the enumeration problem over
words following different formalisms. For example, \cite{bagan2006mso,courcelle2009linear,segoufin2013enumerating} studied the enumeration problem for MSO logic,~\cite{FlorenzanoRUVV20,amarilli2018constant} for regular spanners (i.e. automata), and~\cite{grez2019} for streaming evaluation in complex event processing. For all these formalisms, it is shown that there exists an enumeration algorithm with linear time preprocessing and delay constant in the size of the input word. 

The interest of an efficient enumeration algorithm is to provide a process that can quickly give the firsts answers. Unfortunately, these answers may not be relevant for the user; that is, the enumeration process does not assume how the output will be ordered. A classical manner of considering the user's preferences is to associate a score to each solution and then rank them following this score. This approach has been used particularly in the context of information extraction. Indeed, there have been several recent proposals~\cite{DoleschalKMP20,BennyArxiv} to extend document spanners with annotations from a semi-ring. The proposed annotations are typically useful to capture the confidence of each solution~\cite{DoleschalKMP20}.  For instance, \cite{BennyArxiv} proves that the enumeration of the answers following their scores' order is possible with polynomial-time preprocessing and polynomial delay.

In this paper, we are interested in establishing a framework for scoring outputs and improve the bounds proved in \cite{DoleschalKMP20}. We propose using what we called MSO cost functions, which are formulas in weighted logics~\cite{DrosteG05} extended with open variables. These formulas provide a simple formalism for defining the output and scoring with MSO logic. We show that one can translate each MSO cost function to a cost transducer. These machines are a restricted form of weighted functional vset-automaton~\cite{DoleschalKMP20}, for which there exists at most one run for any word and any valuation. 
We use cost transducers to study the ranked enumeration problem: enumerate all outputs in increasing rank order. Specifically, the main result of the paper is an algorithm for enumerating all the solutions of a cost transducer in increasing order efficiently; specifically, with a linear preprocessing phase and a logarithmic delay between solutions.

Our approach generalizes an algorithm for enumerating solutions proposed in~\cite{grez2019,FlorenzanoRUVV20}.  The preprocessing part builds a heap containing the answers with their score, and one step of the enumeration is simply a pop of the heap. For this, we use a general data structure that we called Heap of Words (HoW), having the classical heap operations of finding/deleting the minimal element, adding an element, and melding two heaps. We also need to add two new operations that allow us to concatenate a letter to and increase the score of all elements of the heap. Finally, we require that this structure is fully-persistent~\cite{driscoll1986making}, i.e., that each of the previous operations returns a new heap without changing the previous one.  To obtain the required efficiency, we rely on a classical persistent data structure called Brodal queue that we extend in order to capture the new operations over the stored words and scores presented above. We call this extension an incremental Brodal queue.

Finally, for ranked query evaluation, there has been recent progress in the context conjunctive queries: on the efficient computation of top-$k$ queries~\cite{TziavelisGR20} and the efficient ranked enumeration~\cite{TziavelisAGRY20,KoutrisArxiv}. These advances consider relational data (which is more general than words) and conjunctive queries (which is more restricted than MSO queries); they are thus incomparable to our work. However, it is important to note some similarities with our work, such as the need for an ``advanced'' priority queue (the Fibonacci heap~\cite{KoutrisArxiv}), which means that our incremental queues might be of great interest there.
 
\smallskip
\noindent \textbf{Contributions}
The contributions of this paper are threefold: \textbf{(i)} we introduce MSO cost functions, a framework to express MSO queries and scores, generalizing the proposals of document spanners; \textbf{(ii)} we give a ranked enumeration scheme that has linear preprocessing time and logarithmic delay in data complexity with a polynomial combined complexity; \textbf{(iii)} we introduce two new data structures for our scheme: the Heaps of Words and the incremental Brodal queues. Both of these structures might be of interest in other ranked enumerations schemes.

\smallskip
\noindent \textbf{Organization.}
Section~\ref{sec:preliminaries} introduces the ranked enumeration
problem for MSO queries on words. Section~\ref{sec:mso}
presents MSO cost functions which is our framework to rank MSO
queries. Section~\ref{sec:renumalgo} describes our enumeration scheme
that rely on two data structures: the Heap of Words described in
Section~\ref{sec:ipqs} and the incremental Brodal queues presented in
Section~\ref{sec:queues}. We finish with some conclusions in Section~\ref{sec:conclusions}.

\section{Preliminaries} \label{sec:preliminaries}

\noindent\textbf{Words.} We denote by $\Sigma$ a finite alphabet, $\Sigma^*$ all words over $\Sigma$, and $\epsilon$ the empty-word of $0$~length. Give a word $w = a_1 \ldots a_n$, we write $w[i] = a_i$.  For two words $u, v \in \Sigma^*$ we write $u \cdot v$ as the concatenation of $u$ and $v$. We denote by $[n] = \{1, \ldots, n\}$.

\smallskip
\noindent\textbf{Ordered groups.} A group is a pair $(\bbG, \gop, \gid)$ where $\bbG$ is a set of elements, $\gop$ is a binary operation over $\bbG$ that is associative, $\gid \in \bbG$ is a
neutral element for $\gop$ (i.e. $\gid \gop g = g \gop \gid
= g$) and every $g \in \bbG$ has an inverse with respect to $\gop$ (i.e. $g \gop g^{-1} = \gid$ for some $g^{-1} \in \bbG$).
A group is abelian if, in addition, $\gop$ is commutative (i.e. $g_1 \gop g_2 = g_2 \gop g_1$). From now on, we assume that all groups are abelian. 
We say that $(\bbG, \gop, \gid,\preceq)$ is an ordered group if $(\bbG, \gop, \gid)$ is a group and $\preceq$ is a total order over $\bbG$ that respects
$\gop$, namely, if
$g_1 \preceq g_2$ then $g_1 \gop g \preceq g_2 \gop g$ for every $g, g_1, g_2 \in \bbG$. 
Examples of (abelian) ordered groups are
$(\bbZ, +, 0, \leq)$ and $(\bbZ^k, +, (0, \ldots,
0),\leq_k)$ where $\leq_k$ represents the lexicographic order over $\bbZ^k$. 

\smallskip
\noindent\textbf{MSO.} We use monadic second-order logic for defining properties over words. As usual, we encode words as logical structures with an order predicate and unary predicates to represent the order and the letters of each positions of the word, respectively. More formally, fix an alphabet $\Sigma$ and let $w \in \Sigma^*$ be a word of length $n$. We encode $w$ as a structure $([n], \leq, (P_a)_{a \in \Sigma})$ where $[n]$ is the domain, $\leq$ is the total order over $[n]$, and $P_a = \{i \mid w[i] = a\}$. By some abuse of notation, we also use $w$ to denote its corresponding logical structure.

A MSO-formula $\varphi$ over $\Sigma$ is given by:
\[
\begin{array}{rcl}
\varphi & := & x \leq y \ \mid \ P_a(x) \ \mid \ x \in X \ \mid \ \varphi \wedge \varphi \ \mid \ \neg \varphi \ \mid \ \exists x. \, \varphi \ \mid \ \exists X. \, \varphi
 \end{array}
\]
where $a \in \Sigma$, $x$ and $y$ are first-order (FO) variables, and $X$ is a monadic second order (MSO) variable (i.e. a set variable). We write $\varphi(\bar{x}, \bar{X})$ where $\bar{x}$ and $\bar{X}$ are the sets of free FO and MSO variables of $\varphi$, respectively.
An assignment $\sigma$ for $w$ is a function $\sigma: \x \cup \X \rightarrow 2^{[n]}$ such that $|\sigma(x)| = 1$ for every $x \in \bar{x}$ (note that we treat FO variables as a special case of MSO variables).
As usual, we denote by $\dom(\sigma) = \x \cup \X$ the domain of the function $\sigma$.
Then we write $(w, \sigma) \models \varphi$ when $\sigma$ is an assignment over $w$, $\dom(\sigma) = \bar{x} \cup \bar{X}$, and $w$ satisfies $\varphi(\bar{x}, \bar{X})$ when each variable in $\bar{x} \cup \bar{X}$ is instantiated by $\sigma$ (see~\cite{libkin2013elements}).
Given a formula $\varphi(\bar{x}, \bar{X})$, we define $
\sem{\varphi}(w) \ = \ \{\sigma \mid (w, \sigma) \models \varphi(\bar{x}, \bar{X})\}
$.
For the sake of simplification, from now on we will only use $\X$ to denote the free variables of $\varphi(\X)$ and use $X \in \X$ for  an FO or MSO variable.

For any assignment $\sigma$ over $w$, we define the support of $\sigma$, denoted by $\supp(\sigma)$, as the set of positions mentioned in $\sigma$; formally, $\supp(\sigma) = \{i \mid \exists v \in \dom(\sigma). \, i \in \sigma(v)\}$. Furthermore, we encode assignments as sequences over the support as follows. 
Let $\supp(\sigma) = \{i_1, \ldots, i_m\}$ such that $i_j < i_{j+1}$ for every $j < m$. Then we define the (word) encoding of $\sigma$ as:
\[
\enc(\sigma) \ = \ (\X_1, i_1) (\X_2, i_2) \ldots (\X_m, i_m) 
\]
such that $\X_{j} = \{X \in \dom(\sigma) \mid i_j \in \sigma(X)\}$ for every $j \leq m$. That is, we represent $\sigma$ as an increasing sequence of positions, where each position is labeled with the variables of $\sigma$ where it belongs. This is the standard encoding used to represent assignments for running algorithms regarding MSO formulas~\cite{bagan2006mso,courcelle2009linear}. Finally, we define the size of $\sigma$ as $|\enc(\sigma)| = |\dom(\sigma)| \cdot m$.
\smallskip

\noindent \textbf{Enumeration algorithms.} Given a formula $\varphi$ and a word $w$, the main goal of the paper is to study the enumeration of assignments in $\sem{\varphi}(w)$.
We give here a general definition of enumeration algorithm and how we measure its delay. Later we use this to define the ranked enumeration problem of MSO. 

As it is standard in the
literature \cite{bagan2006mso,courcelle2009linear,segoufin2013enumerating},
we consider algorithms on Random Access Machines (RAM) with uniform
cost measure~\cite{aho1974design} equipped with addition and
subtraction as basic operations. A RAM has read-only input registers
(containing the input $I$), read-write work memory registers and
write-only output registers.  We say that an algorithm $\mathcal{E}$
is an enumeration algorithm for MSO evaluation if $\mathcal{E}$ runs
in two phases, for every MSO-formula $\varphi$ and a word $w$.
\begin{enumerate}
	\item The first phase, called the preprocessing phase, does not produce output, but may prepare
	data structures for use in the next phase.
	\item The second phase, called the enumeration phase, occurs immediately after the precomputation
	phase. During this phase, the algorithm:
	\begin{itemize}
		\item writes $\# \enc(\sigma_1) \# \enc(\sigma_2) \# \ldots \# \enc(\sigma_k) \#$ to the output registers where $\#$ is a distinct separator symbol, and $\sigma_1$, $\ldots$, $\sigma_k$ is an enumeration (without repetition) of the assignments of $\sem{\varphi}(w)$;
		\item it writes the first $\#$ as soon as the enumeration phase starts,
		\item it stops immediately after writing the last $\#$.
	\end{itemize}
\end{enumerate}	
The separation of $\mathcal{E}$'s operation into a preprocessing and
enumeration phase is done to be able to make an output-sensitive
analysis of $\mathcal{E}$'s complexity.  Formally, we say that
$\mathcal{E}$ has preprocessing time $f: \bbN^2 \to \bbN$ if there
exists a constant $C$ such that the number of instructions that
$\mathcal{E}$ executes during the preprocessing phase on input
$(\varphi, w)$ is at most $C \times f(|\varphi|, |w|)$ for every
MSO-formula $\varphi$ and word $w$.  Furthermore, we measure the delay
as follows. Let $\text{time} _i(\varphi , w)$ denote the time in the
enumeration phase when the algorithm writes the $i$-th $\#$ (if it
exists) when running on input $(\varphi, w)$. Define
$\text{delay}_i(\varphi, w) = \text{time}_{i+1}(\varphi,
w)-\text{time}_{i}(\varphi, w)$.  Further, let
$\text{output}_i(\varphi , w)$ denote the $i$-th element that is
output by $\mathcal{E}$ when running on input $(\varphi, w)$, if it
exists.  We say that $\mathcal{E}$ has delay $g: \bbN^2 \to \bbN$ if
there exists a constant $D$ such that, for all $\varphi$ and $w$, it
holds that:
\[
\text{delay}_i(\varphi , w) \ \leq \ D \times
|\text{output}_i(\varphi, w)| \times g(|\varphi|, |w|)
\]
for every
$i \leq |\sem{\varphi}(w)|$.  Furthermore, if $\sem{\varphi}(w)$ is
empty, then $\text{delay}_1(\varphi,w) \leq k$, namely, it ends in
constant time.  Finally, we say that $\mathcal{E}$ has preprocessing
time $f: \bbN \to \bbN$ and delay $g: \bbN \to \bbN$ in \emph{data
complexity}, if there exists a function $c: \bbN \to \bbN$ such that $\mathcal{E}$ has
preprocessing time $c(|\varphi|) \times f(|w|)$ and
has delay $c(|\varphi|)\times g(|w|)$ (i.e., $f$ and
$g$ describe the complexity once $\varphi$ is considered as fixed).

It is important to notice that, although we fix a particular encoding for assignments and we restrict the enumeration algorithms to this encoding, we can use any encoding for the assignments whenever there exists a linear transformation between $\enc(\cdot)$ and the new encoding. Given the definition of delay, if we use an encoding $\enc'(\sigma)$ for $\sigma$, and there exists a linear time transformation between $\enc(\sigma)$ and $\enc'(\sigma)$ for every $\sigma$, then the same enumeration algorithm works for $\enc'(\cdot)$. In particular, whenever the encoding depends linearly over $\supp(\sigma)$ and $|\bar{x} \cup \bar{X}|$, then the aforementioned property holds. 

\smallskip
\noindent \textbf{Ranked enumeration.} For an MSO formula $\varphi$ and $w \in \Sigma^*$, we consider the ranked enumeration of the set $\sem{\varphi}(w)$. For this, we need to assign an order to the outputs and we do this by mapping each element to a total order set. 
Fix a set $C$ with a total order $\preceq$ over~$C$. A cost function is any partial function $\costf$ that maps words $w \in \Sigma^*$ and assignments $\sigma$ to elements in $C$. Without loss of generality, we assume that $\costf$ is defined only over pairs $(w, \sigma)$ such that $\sigma$ is an assignment over $w$.

Let $\varphi$ be an MSO formula and $\costf$ a cost function over $(C, \preceq)$. We define the ranked enumeration problem of $(\varphi, \costf)$ as
\vspace{-.2cm}
\begin{center}
	\framebox{
		\begin{tabular}{rl}
			\textbf{Problem:} & $\RENUM[\varphi, \costf]$\\
			\textbf{Input:} & A word $w \in \Sigma^*$.  \\
			\textbf{Output:} & Enumerate all $\sigma_1, \ldots, \sigma_k \in \sem{\varphi}(w)$ without \\
			& repetitions  and such that $\costf(w, \sigma_i) \preceq \costf(w, \sigma_{i+1})$. 
		\end{tabular}
	}
\end{center}
Note that we consider the version of the problem in data-complexity where $\varphi$ and $\costf$ are fixed. 
We say that $\RENUM[\varphi, \costf]$ can be solved with preprocessing time $f(n)$ and delay $g(n)$ if there exists an enumeration algorithm $\mathcal{E}$ that runs with preprocessing time $f(n)$ and delay $g(n)$ and, for every $w \in \Sigma^*$, $\mathcal{E}$ enumerates $\sem{\varphi}(w)$ in increasing ordered according to $\costf$. In the next section, we give a language to define cost functions and we state our main result.

\section{MSO cost functions} \label{sec:mso}

To state our main result about ranked enumeration of MSO, first we need to choose a formalism to define cost functions. We do this by staying in the same setting of MSO logic by considering weighted logics over words~\cite{DrosteG05,droste2009handbook,KreutzerR13}. 
Functions defined by extensions of MSO has been studied by using weighted automata, but also people have found it counterparts by extending MSO with a semiring. We use here a fragment of weighted MSO parametrized by an ordered group to fit our purpose.

Fix an ordered group $(\bbG, \gop, \gid, \preceq)$. A weighted MSO-formula $\alpha$ over $\Sigma$ and $\bbG$ is given by the following syntax:
\[
\begin{array}{rcl}
\alpha & := & [\varphi \mapsto g] \ \mid \ \alpha \gop \alpha \ \mid \ \gsum x. \, \alpha
\end{array}
\]
where $\varphi$ is an MSO-formula, $g \in \bbG$, and $x$ is an FO variable. Further,  we  assume that the $\gsum x$ quantifier cannot be nested. For example, $(\gsum x. [\varphi \mapsto g]) \oplus (\gsum y. [\varphi' \mapsto g'])$ is a valid formula but $\gsum x. \gsum y. [\varphi \mapsto g]$ is not. Similar than for MSO formulas, we write $\alpha(\bar{x}, \bar{X})$ to state explicitly the sets of FO-variables $\bar{x}$ and of MSO variables $\bar{X}$ that are free in $\alpha$. 

Let $\sigma$ be an assignment. For any FO-variable $x$ and $i \in \bbN$ we denote by $\sigma[x \rightarrow i]$ the extension of $\sigma$ with $x$ assigned to $i$, namely, $\dom(\sigma[x \rightarrow i]) = \{x\} \cup \dom(\sigma)$ such that $\sigma[x \rightarrow i](x) = \{i\}$ and $\sigma[x \rightarrow i](y) = \sigma(y)$ for every $y \in \dom(\sigma) \setminus \{x\}$. We define the semantics of a weighted MSO formula $\alpha$ as a function from words and assignments to elements in $\bbG$. Formally, for every $w \in \Sigma$ and every assignment $\sigma$ over $w$ we define the output $\sem{\alpha}(w, \sigma)$ recursively as follows:
\[
\begin{array}[t]{ll}
\begin{array}{l}
\sem{[\varphi \mapsto g]}(w, \sigma) = \left\{ 
\begin{array}{ll}
g & (w, \sigma) \models \varphi \\
\gid & \text{otherwise}.
\end{array}
\right. 
\end{array} \ \ \ \ \  & 
\begin{array}{rcl}
\sem{\alpha \gop \alpha'}(w, \sigma) & \!\!\!\! = \!\!\!\! & \sem{\alpha}(w, \sigma)  \gop \sem{\alpha'}(w, \sigma) \vspace{4mm} \\
\sem{\gsum x. \, \alpha}(w, \sigma) &  \!\!\!\! = \!\!\!\! & \displaystyle \biggop_{i=1}^{|w|} \ \sem{\alpha}(w, \sigma[x \rightarrow i])
\end{array}
\end{array}
\]
where $\varphi$ is any MSO-formula, $\alpha$ and $\alpha'$ are weighted MSO formulas, and $g \in \bbG$. By some abuse of notation, in the following we will not make distinction between $\alpha$ and $\sem{\alpha}$, that is, the cost function over $\bbG$ defined by $\alpha$. 
\begin{example}
	Consider the alphabet $\{a,b\}$ and suppose that we want to define a cost function that counts the number of $a$-letters between two variables $x$ and $y$. This can be defined in weighted MSO over $\bbZ$ as follows:
	\[
	\alpha_1 \ := \ \gsum z. \big[(x \leq z \wedge z \leq y \wedge P_a(z)) \mapsto 1 \big]
	\]
	Here, $\alpha_1$ use $z$ to count over all positions of the word and we count~$1$ whenever $z$ is labeled with $a$ and is between $x$ and $y$, and we count $0$, otherwise, which is the identity of $\bbZ$. 
\end{example}
\begin{example}
	Consider again the alphabet $\{a,b\}$ and suppose that we want a cost function to compare assignments over variables $(x, y)$ lexicographically. For this, we can write a weighted MSO-formula over $\bbZ^2$ that maps each assignment $\sigma$ over $x$ and $y$ to a pair $(\sigma(x), \sigma(y))$.  This can be defined in weighted MSO over $\bbZ^2$ as follows:
	\[
\alpha_2 \ := \ \left(\gsum z_1. \, \big[(z_1 \leq x) \mapsto (1,0) \big] \right) + \left(\gsum z_2. \, \big[(z_2 \leq y) \mapsto (0,1) \big] \right)
	\]
	Similar than for the previous example, we use the $\gsum$-quantifier to add in the first and second component the value of $x$ and $y$, respectively. In fact, for every assignment $\sigma = \{x \rightarrow i, y \rightarrow j\}$ over $w \in \Sigma^*$ it holds that $\sem{\alpha_2}(w, \sigma) = (i, j)$. 
\end{example}

Strictly speaking, the syntax and semantics of weighted MSO defined above is a restricted version of weighted logics~\cite{DrosteG05}, in the sense that weighted logics is usually defined over a semiring, which has two binary operations $\gop$ and $\sop$. Although it will be interesting to extend our results for weighted logics over semiring, we leave this for future work. 

We are ready to state the main result of the paper about ranked enumeration of MSO. 
\begin{theorem}
	Fix an alphabet $\Sigma$ and an ordered group $\bbG$. For every MSO formula $\varphi$ over~$\Sigma$ and every weighted MSO formula $\alpha$ over~$\Sigma$ and $\bbG$ the problem $\RENUM[\varphi, \alpha]$ can be solved with linear preprocessing time and logarithmic delay. 
\end{theorem}   
We show applications of this result in the framework of document spanners~\cite{FaginKRV15,DoleschalKMP20}  and the setting of complex event processing~\cite{grez2019}. Due to space restrictions, we address these applications in the appendix.

As it is common for MSO logic over words, we prove this result by developing an enumeration algorithm using automata theory. Specifically, we define a weighted automata model, that we called cost transducer, and show that its expressibility is equivalent to the combination of (boolean) MSO and weighted MSO logic.

From now on, fix an input alphabet $\Sigma$ and an output alphabet~$\Gamma$. Furthermore, fix an ordered group $(\bbG, \gop, \gid, \preceq)$.
A \emph{cost transducer} over $\bbG$ is a tuple $\cT = (Q, \Delta,\costf,I,F)$, where $Q$ is the set of states, $\Delta \subseteq Q \times \Sigma \times 2^{\Gamma} \times Q$ is the transition relation, $\costf: \Delta \rightarrow \bbG$ is a function that associates a cost to every transition of $\Delta$, and $I: Q \rightarrow \bbG$, $F:Q \rightarrow \bbG$ are partial functions that associate a cost in $\bbG$ to (some) states in $Q$. 
The functions $I$ and $F$ are partial functions because they naturally define the set of initial and final states as $\dom(I)$ and $\dom(F)$, respectively.
A run of $\cT$ over a word $w = a_1 a_2 \ldots a_n$ is a sequence of transitions $\rho: q_0 \trans{a_1 / \X_1} q_1 \trans{a_2/ \X_2} \ldots \trans{a_n/ \X_n} q_n$ such that $q_0 \in \dom(I)$ and $(q_{i-1},a_i,\X_i,q_i) \in \Delta$ for every $i \leq n$.
We say that $\rho$ is accepting if $q_n \in \dom(F)$.

For every accepting run $\rho$ as defined above, let $\{i_1, \ldots, i_m\} \subseteq [n]$ be all the positions of $\rho$ such that $\X_{i_j} \neq \emptyset$ and $i_{j} < i_{j+1}$ for all $j \leq m$.
Then we define the output of $\rho$ as the sequence: 
\[
\out(\rho) \ = \ (\X_{i_1},i_1) (\X_{i_2},i_2) \ldots (\X_{i_m},i_m)
\]
Moreover, we extend $\costf$ over accepting runs $\rho$ by adding the costs of all transitions of $\rho$ plus the initial and final cost, namely:
\[
\costf(\rho) = I(q_0) \gop \biggop_{i=1}^{|w|} \costf((q_{i-1},a_i,\X_i,q_i)) \gop F(q_n).
\]
Note that $\out(\rho)$ defines the encoding of some assignment $\sigma$ over $w$ with $\dom(\sigma) = \Gamma$ and $\out(\rho) = \enc(\sigma)$.
Of course, the opposite direction is not true: for some assignment $\sigma$ there could be no run $\rho$ that defines $\sigma$ and, moreover, there could be two runs $\rho_1$ and $\rho_2$ such that $\out(\rho_1) = \out(\rho_2) = \enc(\sigma)$, but $\costf(\rho_1) \neq \costf(\rho_2)$. 
For this reason, we impose an additional restriction to cost transducers: we assume that all cost transducers in this paper are unambiguous, that is, for every $w \in \Sigma^*$ there does not exist two runs $\rho_1$ and $\rho_2$ of $w$ such that $\out(\rho_1) = \out(\rho_2)$.
In other words, a cost transducers satisfies that for every $w \in \Sigma^*$ and assignment $\sigma$ there exists at most one run $\rho$ such that $\out(\rho) = \enc(\sigma)$.

Given the unambiguous restriction of cost transducers, we can define a partial function from pairs $(w, \sigma)$ to $\bbG$ as $\costt{\cT}(w, \sigma) = \costf(\rho)$ whenever there exists a run $\rho$ of $w$ such that $\out(\rho) = \enc(\sigma)$. Otherwise $\costt{\cT}(w, \sigma)$ is not defined. 
Given that for some pairs $(w, \sigma)$ the function $\costt{\cT}$ is not defined, we can define the set $\sem{\cT}(w) = \{\sigma \mid \costt{\cT}(w, \sigma) \text{ is defined}\}$ of all outputs of $\cT$ over $w$.

It is important to notice that, given $w \in \Sigma^*$, a cost transducer $\cT$ is in charged of (1) defining the set of assignments $\sem{\cT}(w)$ and (2) assigning a cost $\sem{\cT}(w, \sigma)$ for each output $\sigma \in \sem{\cT}(w)$. These two task are separated in our setting of ranked MSO enumeration by having a MSO formula $\varphi$ that defines the outputs $\sem{\varphi}$ and a weighted MSO formula $\alpha$ to assign a cost to each pair $(w, \sigma)$. In fact, one can show that cost transducers are equally expressive than combining MSO plus weighted MSO (see proof in appendix \ref{apx:proofOfWMSO-CT}): 
\begin{proposition} \label{prop:WMSO-CT}
	For every cost transducer $\cT$, there exists a MSO formula $\varphi_{\cT}$ and weighted MSO formula $\alpha_{\cT}$ such that $\sem{\cT} = \sem{\varphi_{\cT}}$ and $\costt{\cT}(w,\sigma) = \sem{\alpha_{\cT}}(w,\sigma)$ for every $\sigma \in \sem{\cT}(w)$. Moreover, for every MSO formula $\varphi$ and weighted MSO formula $\alpha$, there exists a cost transducer $\cT_{\varphi, \alpha}$ such that $\sem{\varphi} = \sem{\cT_{\varphi, \alpha}}$ and $\sem{\alpha}(w,\sigma) = \costt{\cT_{\varphi, \alpha}}(w,\sigma)$ for every $\sigma \in \sem{\varphi}(w)$.
\end{proposition}

By the previous result, we can represent pairs of formulas $(\varphi, \alpha)$ by using cost transducers and vice-versa. Similar than for MSO~\cite{reinhardt2002complexity}, there exists a non-elementary blow-up for going from $(\varphi, \alpha)$ to a cost transducer and this blow-up cannot be avoided~\cite{FrickG04}. 

To solve the problem $\RENUM[\varphi, \alpha]$ we can use a cost transducer $\cT_{\varphi, \alpha}$ to enumerate all its outputs following the cost assigned by this machine. More concretely, we study the following rank enumeration problem for cost transducers:
\vspace{.3cm}
\begin{center}
	\framebox{
		\begin{tabular}{rl}
			\textbf{Problem:} & $\RENUMT$\\
			\textbf{Input:} & A cost transducer $\cT$ and a word $w \in \Sigma^*$.  \\
			\textbf{Output:} & Enumerate all $\sigma_1, \ldots, \sigma_k \in \sem{\cT}(w)$  without repetitions and \\ 
			& such that  $\costt{\cT}(w, \sigma_i) \preceq \costt{\cT}(w, \sigma_{i+1})$. 
		\end{tabular}
	}
\end{center}
\vspace{.3cm}
Note that for $\RENUMT$ we consider the cost transducer as part of the input\footnote{In Section~\ref{sec:preliminaries} we introduce the setting of ranked enumeration for MSO formulas and cost functions. One can easily extend this setting and  the definiton of enumeration algorithms for cost transducer.}.
Indeed, for this case we can provide an enumeration algorithm with stronger guarantees regarding the preprocessing time in terms of $\cT$. We now give the theorem formalizing the main result of this paper, which will be proven in the next section:

\begin{theorem}\label{theo:renumt}
	The problem $\RENUMT$ can be solved with $|\cT|\cdot |w|$
	preprocessing time and $\log(|\cT|\cdot |w|)$-delay.
\end{theorem} 

In the rest of the paper, we present the above mentioned ranked enumeration algorithm. We start by showing a general algorithm based on a novel data structure called a Heap of Words. In Section~\ref{sec:ipqs}, we provide the implementation of this structure. In Section~\ref{sec:queues}, we show how to implement the incremental Brodal queues, a technical data structure needed to obtain the required efficiency.

\section{Ranked enumeration algorithm} \label{sec:renumalgo}

In this section, we will see how novel data structures can solve the
ranked enumeration problem for cost transducers on words.
We provide an algorithm for the $\RENUMT$ problem, which uses a structure called \emph{\hnamebigcaps} (\hnameshort) as a black box.
We specify the interface of the \hnameshort{}, to then present the full algorithm.
The \hnameshort{} structure is addressed in detail in the next section.
This structure has the property of being fully-persistent. Given that this is a crucial property, we start with a brief introduction to this concept.

\smallskip
\noindent \textbf{Fully-persistent data structures.} 
A data structure is
said \emph{fully-persistent}~\cite{driscoll1986making} when no
operation can modify the data structure. In a fully-persistent data
structure, all the operations return new data structures, without
changing the original ones. While this seems to be a restriction on
the possible operations, it allows ``sharing''.  For instance, with
a fully-persistent linked list data structure, we can keep two lists
$l_1, l_2$ with $l_1$ being some value followed by the content of
$l_2$ and since no operation modifies the content of $l_1$ or $l_2$
there is no risk that an access to $l_1$ modifies indirectly $l_2$.
In contrast, if we had allowed an operation that modifies the first
value of a list in place (i.e. without returning a new list containing
the modification), the applying this new operation on $l_2$ would have
modified both $l_1$ and $l_2$.

All data structures that we use in this paper are fully-persistent. We
use these data structures to store and enumerate the outputs of the
cost transducer while, at the same time, share and modify the outputs
without any risk of losing them. For more information of
fully-persistent data structures, we refer the reader
to~\cite{driscoll1986making}.

\smallskip
\noindent \textbf{The \hnameshort{} data structure.} 
A \hnamebigcaps{} (\hnameshort{}) over an ordered group $(\bbG, \gop, \gid,\preceq)$ is a data structure $\hstruct$ that stores a finite set $\{\hpair{w_1}{g_1}, \ldots, \hpair{w_n}{g_n}\}$ where each $\hpair{w_i}{g_i}$ is a pair composed by a word $w_i \in \Sigma^*$ and a priority $g_i \in \bbG$. 
Further, we assume that $w_i \neq w_j$ for every $i \neq j$, namely, the stored words form a set too. 
We define $\sem{h} = \{w_1, \ldots, w_n\}$ as the content of $h$ and, given the previous restriction, there is a one-to-one correspondence between $\hpair{w_i}{g_i}$ and $w_i$.
Notice that we will usually write $\hstruct = \{\hpair{w_1}{g_1}, \ldots, \hpair{w_n}{g_n}\}$ to denote that $\hstruct$ stores $\hpair{w_1}{g_1}, \ldots, \hpair{w_n}{g_n}$ but, strictly speaking, $\hstruct$ is a data structure (i.e., a heap). 
Finally, we denote by $\emptyset$ the empty \hnameshort{}.

The purpose of a \hnameshort{} $\hstruct$ is to store words and retrieve quickly the pair $\hpair{w}{g}$ with minimum priority with respect to the order $\preceq$ of the group.  
We also want to manage $\hstruct$ by \emph{deleting} the word with minimum priority, \emph{adding} new words, \emph{increasing} the priority of all elements by some $g \in \bbG$, or \emph{extending} all words with a new letter $a \in \Sigma$. 
Furthermore, we want to build the union of two \hnameshortp{}. More formally, we consider the following set of functions to manage \hnameshortp{}. For \hnameshortp{} $\hstruct$, $\hstruct_1$, and $\hstruct_2$, $w \in \Sigma^*$, $g \in \bbG$, and $a \in \Sigma$ we define:
\[
\begin{array}{ll}
\begin{array}[t]{rcl}
w' & := & \hfindmin(\hstruct) \\
\hstruct' & := & \hdeletemin(\hstruct) \\
\hstruct' & := & \hincreaseby(\hstruct, g) \\
\end{array} \ \ \  & \ \ \
\begin{array}[t]{rcll}
\hstruct' & := & \hmeld(\hstruct_1, \hstruct_2) & \text{s.t. $\sem{\hstruct_1} \cap \sem{\hstruct_2} = \emptyset$} \\
\hstruct' & := & \hadd(\hstruct, \hpair{w}{g}) & \text{s.t. $w \notin \sem{\hstruct}$} \\
\hstruct' & := & \hextendby(\hstruct, a)
\end{array}
\end{array}
\]
where $\hstruct'$ is a new \hnameshort{} and $w' \in \Sigma^*$. In general, each of such functions receives a \hnameshort{} and outputs a \hnameshort{} $\hstruct'$.
As it was explained before, this data structure is fully-persistent and, therefore, after applying any of this function, both the output $\hstruct'$ and its previous version $\hstruct$ are available.
Now, we define the semantics of each operation. Let $\hstruct = \{\hpair{w_1}{g_1}, \ldots, \hpair{w_n}{g_n}\}$. The $\hfindmin$ of $\hstruct$ returns a word $w'$ such that $\qpair{w'}{g'}$ is
stored in $\hstruct$ and $g'$ is minimal among all the
priorities stored, formally, $\hpair{w'}{g'} \in \hstruct$ and $g' = \min\{g \mid \hpair{w}{g} \in \hstruct\}$. If there are several $w'$ satisfying this property, one is picked arbitrarily.
Operation $\hdeletemin$ returns a new \hnameshort{} $\hstruct'$ that stores the set represented by $\hstruct$ without the pair of the word returned by $\hfindmin(\hstruct)$, that is, $\hstruct' = \hstruct \setminus \{\hpair{w'}{g'}\}$ where $w' = \hfindmin(\hstruct)$ and $g' = \min\{g \mid \hpair{w}{g} \in \hstruct\}$.
Finally, the functions $\hadd$, $\hincreaseby$, $\hextendby$, and $\hmeld$ are formally defined as:
\newcommand{\xxspace}{\!\!\!\!\!\!}
\[
\renewcommand{\arraystretch}{1.1}
\begin{array}{ll}
\begin{array}{rcl}
\hmeld(\hstruct_1, \hstruct_2)& \xxspace := \xxspace & \hstruct_1 \cup \hstruct_2  \vspace{2mm}\\ 
\hadd(\hstruct, \hpair{w}{g}) & \xxspace := \xxspace & \hstruct \cup \{\hpair{w}{g}\} \ \,  \\
\end{array}
\begin{array}{rcl}
\ \,  \hincreaseby(\hstruct, g) & \xxspace := \xxspace & \{\hpair{w_1}{(g_1 \gop g)}, \ldots, \hpair{w_n}{(g_n \gop g)}\} \vspace{2mm} \\
\hextendby(\hstruct, a) & \xxspace := \xxspace & \{\hpair{(w_1 \cdot a)}{g_1}, \ldots, \hpair{(w_n \cdot a)}{g_n}\}  \\ 
\end{array}
\end{array}
\]
We assume that $\hadd$, $\hincreaseby$, $\hextendby$ and $\hmeld$ take
constant time and $\hfindmin$ takes $\mathcal{O}(|w'|)$ where $w'
= \hfindmin(\hstruct)$. For $\hdeletemin(h)$, if $h$ was built using
$n$ operations $\hadd$, $\hincreaseby$, $\hextendby$ and $\hmeld$
followed by some number of operations $\hdeletemin$ then computing
$\hdeletemin(h)$ takes $\mathcal{O}(|w'| \cdot \log(n))$ where $w'
= \hfindmin(h)$. In the next section we show how to
implement \hnameshortp{} in order to satisfy these requirements. For
now, we assume the existence of this data structure and use it to
solve~\RENUMT.

\begin{algorithm}[t]
	\caption{Preprocessing and enumeration phases for $\RENUMT$.}\label{alg:evaluation}
	\begin{varwidth}[t]{0.5\textwidth}
		\begin{algorithmic}[1]
			\Require $\cT = (Q, \Delta,\costf,I,F)$ and $w = a_1\ldots a_n$.
			\Procedure{Preprocessing}{$\cT,w$}
			\ForEach{$q \in \dom(I)$}
			\State $\hnode_q^0 \gets \hadd(\hempty, \hpair{\epsilon}{I(q)})$
			\EndFor
			\For{$i \textbf{ from } 1 \textbf{ to } n$}
			\ForEach{$t = (p,a_i,\X,q) \in \Delta$}
			\State $\hnode \gets \hnode_p^{i-1}$
			\If{$\X \neq \emptyset$}
			\State $\hnode \gets \hextendby(\hnode, (\X,i))$
			\EndIf
			
			\State $\hnode \gets \hincreaseby(\hnode, \kappa(t))$
			\State $\hnode_q^{i} \gets \hmeld(\hnode_q^{i}, \hnode)$
			\EndFor
			\EndFor
			
			\ForEach{$q \in \dom(F)$}
			\State $\hnode \gets \hincreaseby(\hnode_q^{n}, F(q))$
			\State $\hnode_{\text{out}} \gets \hmeld(\hnode_{\text{out}}, \hnode)$
			\EndFor
			\State \Return $\hnode_{\text{out}}$
			\EndProcedure
		\end{algorithmic}	
	\end{varwidth}
	\qquad\qquad
	\begin{varwidth}[t]{0.4\textwidth}
		\begin{algorithmic}[1]
			\Require A \hname{} $\hnode$.
			\Procedure{Enumeration}{$\hnode$}
			\State $\texttt{write} \ \# $
			\While{$\hnode \neq \hempty$}
			\State $\texttt{write} \ \hfindmin(\hnode)$
			\State $\hnode \gets \hdeletemin(\hnode)$
			\State $\texttt{write} \ \# $
			\EndWhile
			\EndProcedure
		\end{algorithmic}
	\end{varwidth}
\end{algorithm}

\smallskip
\noindent \textbf{The algorithm.}
In Algorithm~\ref{alg:evaluation}, we show the preprocessing phase and the enumeration phase to solve $\RENUMT$. One one hand, the \textsc{Preprocessing} procedure receives a cost transducer $\cT = (Q, \Delta,\costf,I,F)$ and a word $w \in \Sigma^*$, and computes a \hnameshort{} $\hnode_{\text{out}}$. On the other hand, the \textsc{Enumeration} procedure receives a \hnameshort{} (i.e., $\hnode_{\text{out}}$) and enumerates $\enc(\sigma_1), \ldots, \enc(\sigma_k)$ such that $\{\sigma_1, \ldots, \sigma_k\} = \sem{\cT}(w)$ and $\costt{\cT}(w, \sigma_i) \preceq \costt{\cT}(w, \sigma_{i+1})$. 

In both procedures we use \hnameshort{} to compute the set of answers. Indeed, for each $q \in Q$ and each $i \in \{0,\ldots,|w|\}$ we compute a \hnameshort{} $\hnode_q^i$, and also compute a $\hnode_{\text{out}}$ to store the final results. We assume that all \hnameshortp{} are empty (i.e., $\hnode_{\text{out}} = \emptyset$ and $\hnode_q^i = \emptyset$) when the algorithm starts. For each $i$, we call the set $\{\hnode_q^i \mid q \in Q\}$ the $i$-level of \hnameshort{}. Starting from the $0$-level (lines~2-3), the preprocessing phase goes level by level, updating the $i$-level with the previous $(i-1)$-level (lines 4-10).
It is important to note here that the $\hmeld(\hnode_q^{i}, \hnode)$ call (line 10) is well-defined since $\cT$ is unambiguous (i.e. $\sem{\hnode_q^{i}} \cap \sem{\hnode} = \emptyset$).
After reaching the last $n$-level, the algorithm joins all \hnameshortp{} $\{\hnode_q^{n} \mid q\in \dom(F)\}$ into $\hnode_{\text{out}}$, by incrementing first their cost with~$F(q)$ and melding them into $\hnode_{\text{out}}$ (lines 11-13). Finally, the preprocessing phase return $\hnode_{\text{out}}$ as output (line 14).

In order to understand the preprocessing algorithm, one has to notice that all the evaluation is based on a very simple fact. Let $w_i = a_1 \ldots a_i$ and define the set $\Run_\cT(q, w_i)$ of all partial runs of $\cT$ over $w_i$ that end in state $q$. 
For any of such runs $\rho = q_0 \trans{a_1 / \X_1} \ldots \trans{a_i/ \X_i} q_i \in \Run_\cT(q, w_i)$, define the partial cost of $\rho$ as 
$
\costf^*(\rho) = I(q_0) \gop \biggop_{j=1}^{i} \costf((q_{j-1},a_j,\X_j,q_j)) 
$.
After executing \textsc{Preprocessing}, it will hold that:
$\hnode_q^i \ = \ \big\{ \, \hpair{\out(\rho)}{\costf^*(\rho)} \ \mid \ \rho \in \Run_\cT(q, w_i) \, \big\}$.
This is certainly true for $\hnode_q^0$ after lines 2-3 are executed. Then, if this is true for $(i-1)$-level, after the $i$-th iteration of lines 5-10 we will have that $\hnode_q^i$ contains all pairs of the form $\hpair{\out(\rho) \cdot (\X, i)}{\costf^*(\rho) \gop \costf(t)}$ for each $t = (p,a_i,\X,q) \in \Delta$, plus all pairs $\hpair{\out(\rho)}{\costf^*(\rho) \gop \costf(t)}$ for each $t = (p,a_i,\emptyset,q) \in \Delta$ and $\rho \in \Run_\cT(p, w_{i-1})$.
Given that each line takes constant time, we can conclude that the preprocessing phase takes time $\mathcal{O}(|\cT| \cdot |w|)$ as expected.

For the enumeration phase, we extract each output from $\hnode_{\text{out}}$, one by one, by alternating between the $\hfindmin$ and $\hdeletemin$ procedures. Since with $\hdeletemin$ we remove the minimum element of $\hnode$ after printing it, the correctness of the enumeration phase is straightforward.
Notice that this enumeration will print all outputs in increasing order of priority. 
Furthermore, it will not print any output twice given that $\hnode_{\text{out}}$ contains no repetitions.
To bound the time, notice that the number of $\hadd$, $\hincreaseby$, $\hextendby$ and $\hmeld$ functions used during the pre-processing is at most $\mathcal{O}(|\cT| \cdot |w|)$. For this reason, the delay between each output $w'$ is bounded by $\mathcal{O}(\log(|\cT| \cdot |w|) \cdot |w'|)$, satisfying the promised delay between outputs. 

We want to finish this section by emphasizing that the ranked enumeration problem of cost transducers reduces to computing efficiently the \hnameshort{}'s methods. Moreover, it is crucial in this algorithm that this data structure is fully-persistent, and each operation takes constant time. Indeed, this allows us to pass the outputs between levels very efficiently and without losing the outputs of the previous levels.

\section{The implementation of \hnameshort{} data structure} \label{sec:ipqs}

In this section we focus on the \hnameshort{} data structure and explain its implementation using yet another structure called incremental Brodal queue.
We begin by explaining the general technique we use to store sets of strings with priorities, and end by giving a full implementation of the functions to manage \hnameshortp{}.

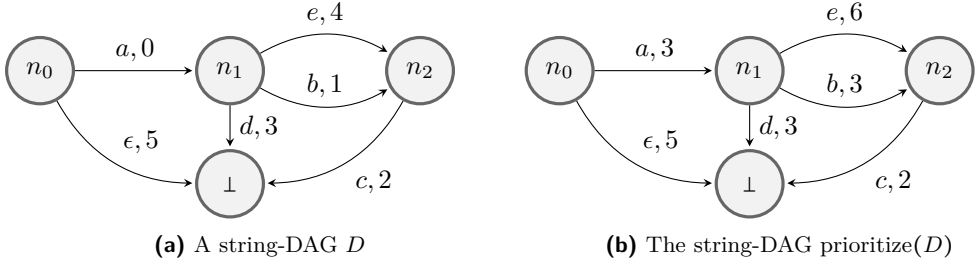
\begin{figure}[t]
	\captionsetup[subfigure]{justification=centering}
	\centering
	\begin{subfigure}[b]{0.48\linewidth}
		\begin{tikzpicture}
		[>=stealth,
		shorten >=1pt,
		node distance=2.5cm,
		on grid,
		auto,
		every state/.style={draw=black!60, fill=black!5, very thick}
		]
		\node[state] 						(0)		{$n_0$};
		\node[state] [right =  of 0]	(1)		{$n_1$};
		\node[state] [right =  of 1]	(2)		{$n_2$};
		\node[state] [below = 1.5cm  of 1]	(3)		{$\bot$};
		
		\path[->]
		
		(0)		edge[]	node			{$a,0$}			(1)
		edge[bend right] node	{$\epsilon,5$}	(3)
		(1)		edge[bend left]	node	{$e,4$}			(2)
		edge[bend right]node	{$b,1$}			(2)
		edge[]node		{$d,3$}			(3)
		(2)		edge[bend left]		node		{$c,2$}			(3)
		;
		\end{tikzpicture}
		\caption{A \sDAG{} $D$} \label{fig:sd}
	\end{subfigure}
	\begin{subfigure}[b]{0.48\linewidth}
		\begin{tikzpicture}
		[>=stealth,
		shorten >=1pt,
		node distance=2.5cm,
		on grid,
		auto,
		every state/.style={draw=black!60, fill=black!5, very thick}
		]
		\node[state] 						(0)		{$n_0$};
		\node[state] [right =  of 0]	(1)		{$n_1$};
		\node[state] [right =  of 1]	(2)		{$n_2$};
		\node[state] [below = 1.5cm  of 1]	(3)		{$\bot$};
		
		\path[->]
		
		(0)		edge[]	node			{$a,3$}			(1)
		edge[bend right]node	{$\epsilon,5$}	(3)
		(1)		edge[bend left]	node	{$e,6$}			(2)
		edge[bend right]node	{$b,3$}			(2)
		edge[]node		{$d,3$}			(3)
		(2)		edge[bend left]		node		{$c,2$}			(3)
		;
		\end{tikzpicture}
		\caption{The \sDAG{} $\fprio(D)$} \label{fig:fsd}
	\end{subfigure}
        \caption{A \sDAG{} $D$ and the result of $\fprio(D)$.}
\end{figure}

Let $\Sigma$ be a possibly infinite alphabet and $\bbG = (\bbG, \gop, \gid,\preceq)$ an order group. A \emph{\sDAG{}} over $\Sigma$ and $\bbG$ is a DAG $D = (V,E)$ where the edges are
annotated with symbols in $\Sigma \cup \{\epsilon\}$ and priorities in $\bbG$. Formally, each edge has the form
$e =(u,a,g,v)$, where $u,v \in V$, $a \in \Sigma \cup \{\epsilon\}$ and~$g \in \bbG$. 
Given a path $\rho = v_1 \xrightarrow{a_1,g_1} \ldots \xrightarrow{a_k,g_k}
v_k$, let $\hpair{w_\rho}{g_\rho}$ be the pair defined by $\rho$, where
$w_\rho = a_1 \ldots a_k$ and $g_\rho = g_1 \gop \ldots \gop g_k$.
We make two more assumptions that any \sDAG{} must satisfy.
First, we assume that there is a special sink vertex
$\bot \in V$ that is reachable from any $v \in V$, has no outgoing
edges, and that all edges with $\epsilon$ must point to $\bot$. 
Second, we assume that, for every $v \in V$ and every two different paths $\rho$ and $\rho'$ from $v$ to $\bot$, it holds that $w_{\rho} \neq w_{\rho'}$.
Given these two assumptions, we say that each $v \in V$ encodes a set of pairs $\sem{D}(v)$: for $v={\bot}$
this set is the empty set, while for all $v\neq \bot$ this set is
defined by all the paths from $v$ to $\bot$, i.e., $\sem{D}(v) = \{
\hpair{w_\rho}{g_\rho} \mid \text{$\rho$ is a path from $v$ to
$\bot$}\}$.  
By these two assumptions, there is a correspondence between the words in $\sem{D}(v)$ and the paths from $v$ to~$\bot$. 
For instance, the strings associated with $n_0$ in
the \sDAG{} depicted in Figure~\ref{fig:sd} are $a d$ with priority
$0+3=3$, $a b c$ with priority $0+1+2=3$, $ a e c $ with priority
$0+4+2=6$ and $\epsilon$ with priority $5$.

This structure is useful to store a big number of strings in a
compressed manner.  Further, since $\epsilon$ can only appear at the
last edge of a path,  by doing a DFS it can be used to retrieve all of them without repetitions and taking
time linear in the length of each string. However, one
can see that it is not very useful when we want to enumerate them by
rank order.  This motivates the following \sDAG{} construction.  We
define a function $\fprio(D)$ that receives a \sDAG{} $D =(V,E)$ and
returns a \sDAG{} $D'=(V,E')$ where each edge $(u,a,g,v)$ of $E$ is
replaced by an edge $(u,a,g\gop g',v)$ in $E'$, where $g'$ is the
minimum priority in $\sem{D}(v)$.  For instance,
Figure~\ref{fig:fsd} shows the \sDAG{} resulting after applying
$\fprio$ to $D$ of Figure~\ref{fig:sd}.  Having $\fprio(D)$ makes
finding the string with minimum priority of a vertex much easier: we simply need
to follow recursively the edge with minimum priority.  In $n_0$ of
Figure~\ref{fig:fsd} we make the path $n_0 \xrightarrow{\!a,3}
n_1 \xrightarrow{b,3} n_2 \xrightarrow{c,2} \bot$ and compute the
minimum pair $[abc,3]$ (the priority is retrieved from the first
edge).

\newcommand{\bQ}{Q}
\newcommand{\bR}{R}

Before presenting the \hnameshort{} implementation, we need to introduce another fully-persistent data structure.
This structure is based on the Brodal queue~\cite{brodal1996optimal}, a known worst-case efficient priority queue, which we extend with the new function $\mincreaseby$. 
Formally, an \emph{incremental Brodal queue}, or just a \emph{queue}, is a fully-persistent data structure $\bQ$ which stores a set $P = \{ \qpair{\mathcal{E}_1}{g_1} \ldots \qpair{\mathcal{E}_k}{g_k} \}$, where each $\mathcal{E}_i$ is a stored element and $g_i$ is its priority.
As an abuse of notation, we often write $\bQ = P$.
The functions to manage incremental Brodal queues include all functions for \hnameshort{} except $\hextendby$, namely $\mfindmin$, $\mdeletemin$, $\madd$, $\mincreaseby$ and $\mmeld$; their definition also remains the same as for \hnameshort{}. Note that we use different fonts to distinguish the operations over \hnameshortp{} versus the operations over incremental Brodal queues. For example, we write $\hfindmin$ for \hnameshortp{}  and $\mfindmin$ for queues.  
Further, this queue has two additional functions: $\misempty$, that checks if the queue is the empty queue~$\emptyset$; and $\mminPriority$, that returns the value $g$ of the minimal priority among all the priorities stored.
For the rest of this section we assume the existence of an incremental Brodal queue structure such that all functions run in time $\cO(1)$ except for $\mdeletemin$, which runs in $\cO(\log(n))$, where $n$ is the number of pairs stored in the queue.
Finally, all these operations are fully-persistent.
The in-detail explanation of this structure is derived to the next section.

With the previous intuition and the structure above, we can now present the implementation for \hnamebigcaps{}.
A \hnameshort{} $h$ is implemented as an incremental Brodal queue $\bQ$ that stores a set $\{ \qpair{(a_1,h_1)}{g_1}, \ldots, \qpair{(a_k,h_k)}{g_k} \}$, where each $a_i \in \Sigma \cup \{\epsilon\}$, each $h_i$ is a \hnameshort{} and each $g_k \in \bbG$.
We write $h = \hq{\bQ}$ to make clear that we are talking about a  \hnameshort{} and not the queue.
The empty \hnameshort{} is simply the empty queue $\hq{\emptyset}$.
Intuitively, the recursive references to \hnameshortp{} are used to encode a \sDAG{} $D$; more specifically, we use it to encode $\fprio(D) = (V,E)$ and store the edges using the queue structure.
For every $u \in V$, we define a \hnameshort{} $h_u = \hq{\bQ}$ such that each pair $\qpair{(a,h_v)}{g}$ stored in $\bQ$ represents an edge $(u,a,g,v) \in E$.
For instance, continuing with the example of Figure~\ref{fig:fsd}, we have a \hnameshort{} for each vertex: $h_\bot = \hq{\emptyset}$, $h_{n_2} = \hq{\{ \qpair{c,h_\bot}{2} \}}$, $h_{n_1} = \hq{\{ \qpair{(e,h_{n_2})}{6},\qpair{(b,h_{n_2})}{3},\qpair{(d,h_\bot)}{3} \}}$ and $h_{n_0} = \hq{\{ \qpair{(a,h_{n_1})}{3},\qpair{(\epsilon,h_\bot)}{5} \}}$.

We now explain the implementation of the functions defined in Section~\ref{sec:renumalgo} to manage~\hnameshort{}.
Consider a \hnameshort{} $h = \hq{\bQ}$.
For each $\textsc{op} \in \{\hmeld, \hincreaseby\}$, the function is just applied directly to the queue, i.e., $\textsc{op}(\hq{\bQ}) = \hq{\op(\bQ)}$.
The implementation of $\hadd$ and the other functions is now described and presented in Algorithm~\ref{alg:how}.

\begin{algorithm}[t]
	\caption{\hnameshort{}'s implementation of $\hadd$, $\hextendby$, $\hfindmin$ and $\hdeletemin$.} \label{alg:how}
	\begin{varwidth}[t]{0.55\textwidth}
		\begin{algorithmic}[1]
			\Procedure{$\hadd$}{$\hq{\bQ},\qpair{a}{g}$}
			\State \Return $\hq{\madd(\bQ,\qpair{(a,\hq{\emptyset})}{g})}$
			\EndProcedure
			
			\smallskip

			\Procedure{$\hextendby$}{$\hq{\bQ},a$}
			\If {$\misempty(\bQ)$}
			\State \Return $\hq{\emptyset}$
			\EndIf
			\State \Return $\hq{\madd(\emptyset,\qpair{(a,\hq{\bQ})}{\mminPriority(\bQ)})}$
			\EndProcedure
			\smallskip
			
			\Procedure{$\hfindmin$}{$\hq{\bQ}$}
			\State  $(a,\hq{\bQ'}) \gets \mfindmin(\bQ)$
			\If{$\misempty(\bQ')$}
			\State \Return $a$
			\EndIf
			\State \Return $\hfindmin(\hq{\bQ'}) \cdot a$
			\EndProcedure
			\algstore{myalg}
		\end{algorithmic}
	\end{varwidth}
	\quad
	\begin{varwidth}[t]{0.45\textwidth}
		\begin{algorithmic}[1]
			\algrestore{myalg}
			\Procedure{$\hdeletemin$}{$\hq{\bQ}$}
			\If {$\misempty(\bQ)$}
			\State \Return $\hq{\emptyset}$
			\EndIf
			\State  $(a,\hq{\bR}) \gets \mfindmin(\bQ)$
			\State  $\bQ' \gets \mdeletemin(\bQ)$
			\State $\hq{\bR'}  \gets \hdeletemin(\hq{\bR})$
			\If{$\iisempty(\bR')$}
			\State \Return $\hq{\bQ'}$
			\EndIf
			\State $\delta \gets \mminPriority(\bR')\gop (\mminPriority(\bR))^{\minus 1}$
			\State $g \gets \mminPriority(\bQ)\gop\delta$
			\State \Return $\hq{\madd(\bQ',\qpair{(a,\hq{\bR'})}{g})}$
			\EndProcedure
		\end{algorithmic}
	\end{varwidth}
\end{algorithm}

In the case of $\hadd(h,a)$, an edge is added that points to $\hq{\emptyset}$; this can be extended to add a word $w$ instead by allowing that edges keep words instead of single letters.
To implement $\hextendby(\hq{\bQ},a)$, we simply need to create a new queue
containing the element $\qpair{(a,\hq{\bQ})}{\mminPriority(\bQ)}$.  For
$\hfindmin(\hq{\bQ})$, to get the minimum element we recursively use
$\mfindmin(\bQ)$ to find the outgoing edge with minimum priority, as we
explained when the $\fprio$ function was introduced.  For
$\hdeletemin$, in order to delete the string with minimum priority, we
use the fact that the set of all paths, minus the one with minimal
priority, is composed by: (1) all the paths that do not start with the
minimal edge, and (2) all the paths starting with the minimal edge
that are followed by any path minus the one with minimal priority.
For instance, in Figure~\ref{fig:fsd}, the minimal path from $n_0$ is
$ \pi = n_0 \xrightarrow{a} n_1 \xrightarrow{d} \bot$.  Then, the set
of paths minus $\pi$ is composed by (1)
$n_0 \xrightarrow{\epsilon} \bot$, and (2) $n_0 \xrightarrow{a}
n_1 \xrightarrow{e} n_2 \xrightarrow{c} \bot$, $n_0 \xrightarrow{a}
n_1 \xrightarrow{b} n_2 \xrightarrow{c} \bot$.  In
procedure \textproc{DeleteMin}, $\hq{\bQ'}$ stores the paths of (1),
while $\hq{\bR'}$ stores the paths from (2) minus the first edge (lines
16-17).  Further, since the minimal path was removed, a new priority
needs to be computed for this edge, which is computed and stored as
$g$ (line~20-21).  This priority is used to create an edge to $\bR'$, i.e. $\qpair{(a,\hq{\bR'})}{g}$, which together represent the paths of (2). 
This is connected with the paths of (1), i.e. $\hq{\bQ'}$, and the result is returned in line~22.
The border case case where (2) is empty is managed by lines~18-19, in which case it simply returns $\hq{\bQ'}$.

We delegate the complexity proofs to the appendix~\ref{apx:how} but
this structures achieves the complexities given in
Section~\ref{sec:renumalgo}.  We end this section by arguing that the
implementation of HoW is fully-persistent. For this, note that the
performance of HoW relies on the implementation of incremental Brodal
queues. Indeed, given that these queues are fully-persistent and each
method in Algorithm~\ref{alg:how} creates new queues without modifying
the previous ones, the whole data structure is
fully-persistent. Therefore, it is left to prove that we can extend
Brodal queues as we already mentioned. We will show this in the next section.

\section{Incremental brodal queues} \label{sec:queues}

In this section, we discuss how to implement an incremental Brodal queue, the last ingredient of our ranked enumeration algorithms for MSO cost functions. This data structure extends Brodal queues~\cite{brodal1996optimal} by including the $\mincreaseby$ procedure. 
Indeed, our construction of incremental Brodal queues follows the same approach as in~\cite{brodal1996optimal}. We start by defining what we call an \emph{incremental binomial heap}, for which most operations take logarithmic time, to then show how to extend it to lower the cost to constant time, except for $\mdeletemin$ that takes logarithmic time. The most relevant aspects for this extension to support $\mincreaseby$ appear in the definition of the incremental binomial heap. For this reason and space restrictions, in this section we present only the implementation of the incremental binomial heap. The details of how to extend it to an incremental Brodal queue can be found in the appendix. We start by introducing some notation to define then the data structure with the operations.

A \emph{multitree structure} is a pair $\M = (\tdom, \fch, \nsb, \troot)$ where $\tdom$ is a set of nodes, $\fch:\tdom \rightarrow \tdom \cup \{\tempty\}$ and $\nsb:\tdom \rightarrow \tdom \cup \{\tempty\}$ are functions such that $\tempty \notin \tdom$ and $\troot \in \tdom$ is a special node.  Further, we assume that the directed graph $G_\M = (\tdom, \{(u, v) \mid \fch(u) = v \text{ or } \nsb(u) = v\})$ is a multitree, namely, it is a directed acyclic graph (DAG) in which the set of vertices reachable from any vertex induces a tree.
Let $V_{\troot}$ denotes the reachable nodes from $\troot$ and $G_{\troot} = (\tdom_{\troot}, \{(u, v) \mid \fch(u) = v \text{ or } \nsb(u) = v\})$ the graph induced by $V_{\troot}$, which is a tree by definition.
Note that $G_{\troot}$ is using the first-child next-sibling encoding to form an ordered forest. To see this, let $\nsb^*(v)$ be the smallest subset of $\tdom$ such that $v \in \nsb^*(v)$ and $\nsb(u) \in \nsb^*(v)$ whenever $u \in \nsb^*(v)$. Then the set $\troots = \nsb^*(\troot)$ represents the roots of the forest and for each $v \in V_{\troot}$ the set $\chd(v) = \nsb^*(\fch(v))$ are the children of the node $v$ in the forest where $\chd(v) = \emptyset$ when $\fch(v) = \tempty$. Here both sets are ordered by the $\nsb$ function, then we will usually write $\troots = v_1, \ldots, v_j$ or $\chd(v) = u_1, \ldots, u_k$ to denote both the elements of the set and its order. 
Also, we write $\parent(v) = u$ if $v \in \chd(u)$ and we say that $v$ is a leaf if $\fch(v) = \bot$.
Note that in $\M$ a node could have different ``parents'' (i.e. $G_\M$ is a DAG) depending on the node $\troot$ that we start. 
We say that $\M$ \emph{forms a tree} if $\nsb(\troot) = \tempty$. 
Furthermore, for $v \in \tdom$ we denote by $\M_v$ the tree hanging from $v$, namely, $\M_v$ is equal to $\M$ with the exception that $\trooti{\M_v} = v$ and $\nsb_{\M_v}(v) = \bot$.
As it will clear below, this encoding will be helpful to build the data structure and assure the persistent requirement. 

A binomial tree of rank $k$ is recursively defined as follows. A binomial tree of rank $0$ is a leaf and a binomial tree of rank $k+1$ is a multitree structure $\M$ that forms a tree such that $\chd(\troot) = u_{k},\ldots, u_0$ and $\M_{u_i}$ is a binomial tree of rank $i$. If $\M$ is a binomial tree we denote its rank by $\trank(\M)$. 
One can easily show by induction over the rank (see~\cite{cormen2009introduction}) that for every binomial tree $\M$ of ranked~$k$, it holds that $|\tdom_{\troot}| = 2^k$ and, thus, the number of children of each node is of logarithmic size with respect to the size of $\T$, i.e., $|\chd(v)| \leq \log(|\tdom_{\troot}|)$ for every $v\in\tdom_{\troot}$. We use this property several times throughout this section.

Fix an ordered group $(\bbG, \gop, \gid,\preceq)$.
An incremental binomial heap over $\bbG$ is defined as a pair
$
\h = (\tdom, \fch, \nsb, \troot, \tdelta, \tval, \tdeltainit)
$
where $(\tdom, \fch, \nsb, \troot)$ is a multitree structure, $\tdelta: \tdom \rightarrow \bbG$ is the delta-priority function, $\tval: \tdom \rightarrow \mathcal{E}$ is the element function where $\mathcal{E}$ is the set of elements that are stored, and $\tdeltainit \in \bbG$ is an initial delta value. Further, if $\M$ is the multitree structure defined by $(\tdom, \fch, \nsb, \troot)$ and $\troots = v_1, \ldots, v_n$ are its roots, then each $\M_{v_i}$ is a binomial tree with $\trank(\M_{v_i}) < \trank(\M_{v_{i+1}})$ for each $i < n$. In other words, an incremental binomial heap has the same underlying structure than a standard binomial heap~\cite{cormen2009introduction}.
Usually in the literature~\cite{brodal1996optimal}, a binomial heap is imposed a min-heap property, meaning that a node always has lower priority than its children, which is crucial for dequeuing elements in order.
Instead, we give to our heap a different semantics by keeping the difference between nodes with the $\tdelta$-function and computing the real priority function $\tprio_{\troot}: \tdom_{\troot} \rightarrow \bbG$ as follows: $\tprio_{\troot}(v) := \tdeltainit \gop \tdelta(v)$ whenever $v$ is a root of the underlying multitree structure, and $\tprio_{\troot}(v) := \tprio_{\troot}(u) \gop \tdelta(v)$ whenever $\parent(v) = u$.
Given that  $\parent(v)$ depends on the starting node $\troot$, then $\tprio_{\troot}$ also depends on $\troot$. 
In addition, we assume that a min-heap property is satisfied over the real priority function, namely,  $\tprio_{\troot}(u) \preceq \tprio_{\troot}(v)$ whenever $\parent(v) = u$. 
Then $\h$ is a heap where each node $v \in \tdom$ keeps a pair $(\tval(v), \tprio_{\troot}(v))$ where $\tval(v)$ is the stored element and $\tprio_{\troot}(v)$ its priority in the heap. This principle of storing the deltas between nodes instead of the real priority is crucial for supporting the increased-by operation of the data structure. 

Next, we show how to implement the operations of an incremental Brodal queue stated in Section~\ref{sec:ipqs}, namely, 
$\misempty$, $\mincreaseby$, $\mfindmin$ $\mminPriority$, $\mdeletemin$, $\madd$, and $\mmeld$.
We implement this with an incremental binomial heap where the only difference is that $\misempty$ and $\mincreaseby$ will take constant time, and $\mfindmin$ $\mminPriority$, $\mdeletemin$, $\madd$, and $\mmeld$ will take logarithmic time.
In the appendix we show how to extend incremental binomial heaps to lower the complexity of $\mfindmin$ $\mminPriority$, $\mdeletemin$,  and $\madd$ to constant time, by using the same techniques as in~\cite{brodal1996optimal}.
Most operations of incremental binomial heaps are similar to the operations on binomial heaps (see~\cite{cormen2009introduction}), however, for the sake of completeness we explain each one in detail, highlighting the main differences to manage the delta priorities.

From now on, fix an incremental binomial heap $\h = (\tdom, \fch, \nsb, \troot, \tdelta, \tval, \tdeltainit)$. 
Given that all operations must be persistent, we will usually create a copy $\h'$ of $\h$ by extending $\h$ with new fresh nodes. More precisely, we will say that $\h'$ is an \emph{extension} of $\h$ (denoted by $\h \subseteq \h'$) iff $\tdomi{\h} \subseteq \tdomi{\h'}$ and $\opi{\h'}(v) = \opi{\h}(v)$ for every $v \in \tdomi{\h}$ and $\op \in \{\fch, \nsb, \tdelta, \tval\}$ (note that $\troot$ and $\tdeltainit$ may change).
Furthermore, for $\h \subseteq \h'$ we will say that a node $v' \in \tdomi{\h'} \setminus \tdomi{\h}$ is a \emph{fresh copy} of $v \in \tdomi{\h}$ if $v'$ in $\h'$ has the same structure as $v$ in $\h$ where only the differences are defined explicitly, namely, we omit the functions that are the same as for $v$. 
For example, if we say that ``$v'$ is a fresh copy of $v$ such that $\nsbi{\h'}(v') := \tempty$'', this means that $\nsbi{\h'}(v') := \bot$ and $\opi{\h'}(v') = \opi{\h}(v)$ for every $\op \neq \nsb$.

The first operation, $\misempty(\h)$, can easily be implemented in constant time, by just checking whether $\troot = \bot$ or not. Similarly, $\mincreaseby(\h,\delta)$ can be implemented in constant time by just updating $\tdeltainit$ to $\tdeltainit \gop \delta$, which is the purpose of having $\tdeltainit$.
For $\mfindmin(\h)$ or $\mminPriority(\h)$, a bit more of work is needed. Recall that a $k$-rank binomial tree with $|\tdom|$ nodes satisfies $|\tdom| = 2^k$. Given that $\troots = v_1, \ldots, v_n$ is a sequence of binomial trees ordered by rank, one can easily see that $n \in \mathcal{O}(\log(|\tdom_{\troot}|)$.
Therefore, we need at most a logarithmic number of steps to find the node $v_i$ with the minimum priority and return $\tval(v_i)$ or $\tprio_{\troot}(v_i)$ whenever $\mfindmin(\h)$ or $\mminPriority(\h)$ is asked, respectively.

For $\madd(\h, e, g)$ or $\mdeletemin(\h)$, we reduce them to melding two heaps. For the first operation, we create a heap $\h'$ whose multitree structure has one node, call it $v$, $\tdeltai{\h'}(v) := g$, $\tvali{\h'}(v) := e$, and $\tdeltainiti{\h'} := \gid$. Then we apply $\mmeld(\h, \h')$ obtaining a heap where the new node $(e, g)$ is added to $\h$. 
For the second operation, we remove the minimum element by creating two heaps and then apply the meld operation. Specifically, let $\troots = v_1, \ldots, v_n$ be the roots of $\h$ and $v_i$ be the root with the minimum priority. Then we build two heaps $\h_1$ and $\h_2$ such that $\h \subseteq \h_{i}$ for $i\in\{1,2\}$. For $\h_1$, we extend $\h$ by creating fresh copies of all $v_j$, $j \neq i$. Formally, define $\tdomi{\h_1} = \tdomi{\h} \cup \{v_1', \ldots, v_n'\}$ where each $v_j'$ is a fresh copy of $v_j$ with the exception of $v_{i-1}'$ that we set $\nsbi{\h_1}(v_{i-1}') := v_{i+1}'$. Finally, define $\trooti{\h_1} = v_1'$ as the starting node of $\h_1$. Now, for $\h_2$ we extend $\h$ by creating a copy of the children of $v_i$ in $\h$ in reverse order and updating $\tdeltainiti{\h}$ to $\tdeltainiti{\h}\gop \tdeltai{\h}(v_i)$ (recall that the children of a binomial tree are ordered by decreasing rank).
Formally, if $\chdi{\h}(v_i) = u_1, \ldots, u_k$, then $\tdomi{\h_2} = \tdomi{\h} \cup \{u_1', \ldots, u_k'\}$ where each $u_j'$ is a fresh copy of $u_j$ such that $\nsbi{\h_2}(u_j') := u_{j-1}'$ for $j > 1$ and $\nsbi{\h_2}(u_1') := \bot$. 
Finally, define $\trooti{\h_2} := u_k'$ and $\tdeltainiti{\h_2} := \tdeltainiti{\h}\gop \tdeltai{\h}(v_i)$.  
The reader can check that $\h_1$ and $\h_2$ are valid incremental binomial heaps and, furthermore, $\h_1$ is $\h$ without $v_i$ and $\h_2$ contains only the children of $v_i$ in reverse order. Therefore, to compute $\mdeletemin(\h)$ we return $\mmeld(\h_1, \h_2)$. Given that the construction of $\h_1$ and $\h_2$ takes at most logarithmic time in the size of $\h$ (i.e. there is at most a log number of roots or children), then the  procedure takes logarithmic time. Furthermore, $\h$ was never touched and then the operation is fully-persistent.

For $\mmeld(\h_1, \h_2)$, we use the same algorithm as for melding two binomial heaps with two modifications that are presented here. For melding two binomial heaps, we point the reader to~\cite{cormen2009introduction} in which this operation is well explained. For the first change, we need to update the link operation~\cite{cormen2009introduction} of two binomial trees to support the use of the delta priorities. Given a incremental binomial heap $\h$ and its underlying multitree structure $\M$, let $v_1$ and $v_2$ be two nodes in $\h$ such that $\tdelta(v_1) \preceq \tdelta(v_2)$ and $\M_{v_1}$ and $\M_{v_2}$ has the same rank $k$. Then the \emph{link} of $v_1$ and $v_2$, denoted by $\mlink(\h, v_1, v_2)$, outputs a pair $(\h', v_1')$ such that $\h'$ is an extension of $\h$ and $\M'_{v_1'}$ is a binomial tree of rank $k+1$ containing the nodes of $\M_{v_1}$ and $\M_{v_2}$. Formally, $\tdomi{\h'} := \tdomi{\h} \cup \{v_1', v_2'\}$ and $v_1'$ and $v_2'$ are fresh copies of $v_1$ and $v_2$ such that $\fchi{\h'}(v_1'):= v_2'$, $\nsbi{\h'}(v_2'):= \fchi{\h}(v_1)$ and $\tdeltai{\h'}(v_2') := \tdeltai{\h}(v_1)^{-1} \gop \tdeltai{\h}(v_2)$. Note that the new node $v_1'$ defines a binomial tree $\M'_{v_1'}$ of rank $k+1$ containing all nodes of $\M_{v_1}$ and $\M_{v_2}$, maintaining the priorities of $\h$ and such that $\tprioi{\h'}(u) \preceq \tprioi{\h'}(u')$ whenever $u = \parent(u')$. The second change of the algorithm in~\cite{cormen2009introduction} is that, before melding $\h_1$ and $\h_2$, we push each initial delta value to the roots of the corresponding data structures. For this, given an incremental binomial heap $\h$ we construct $\h^\downarrow$ with $\h \subseteq \h^\downarrow$ as follows. Let $\trootsi{\h} = v_1, \ldots, v_k$. Then $\tdomi{\h^\downarrow} = \tdomi{\h} \cup \{v_1', \ldots, v_k'\}$ where $v_1', \ldots, v_k'$ are fresh copies of $v_1, \ldots, v_k$ and $\tdeltai{\h^\downarrow}(v_i') := \tdeltainiti{\h} \gop \tdeltai{\h}(v_i)$. Furthermore, we define $\trooti{\h^\downarrow} := v_1'$ and $\tdeltainiti{\h^\downarrow} := \gid$. Note that in $\h^\downarrow$ we can forget about the initial delta value given that this is included in the root of each binomial tree. 
Finally, to meld $\h_1$ and $\h_2$ we construct $\h_1^\downarrow$ and $\h_2^\downarrow$ and then apply the melding algorithm of~\cite{cormen2009introduction} with the updated version of the link function, $\mlink(\h, v_1, v_2)$.  
Overall, the operation takes logarithmic time to build $\h_1^\downarrow$ and $\h_2^\downarrow$, and logarithmic time to meld both heaps. Moreover, given that $\mlink(\h, v_1, v_2)$ and the construction of $\h_1^\downarrow$ and $\h_2^\downarrow$ do not modify the initial heap $\h$, then the meld operation is persistent as well. 

To finish this section, we recall that the next step is to extend the incremental binomial heap to an incremental Brodal queue. For this, we follow the same approach as in~\cite{brodal1996optimal} to lower the time complexity of find-min, add, and meld operation from logarithmic to constant time (see the appendix for further discussion).

\section{Conclusions} \label{sec:conclusions}

This paper presented an algorithm to enumerate the answers of queries over words, in an order defined by a cost function, that has a linear preprocessing and a logarithmic delay in the size of the words. We first introduced the notion of MSO cost functions, to then present a ranked enumeration scheme. This scheme relies on a particular data structure called HoW. The complexity of our algorithms depends mainly on the performance of the operations of HoW. To implement them, we extend a well known persistent data structure called Brodal queue. Thanks to this data structure, we obtain the bounds of our algorithm.

For future work, we would like to find a lower bound that justifies the logarithmic delay or whether one can achieve a better delay.  We also plan to study how the introduced data structures and algorithms could be used in other enumeration schemes (e.g., relational databases). 
Finally, we would also like to validate our approach in practical settings.

\newpage

\onecolumn
\appendix

\section{Applications} \label{sec:applications}

In this section we show the application of our main result in two different settings related to MSO logic over words: document spanners and complex event processing. 

\subsection{Document spanners}


The framework of document spanners was proposed in~\cite{FaginKRV15} as a formalization of ruled-based information extraction and has attracted a lot of attention both in terms of the formalism~\cite{freydenberger2018joining, maturana2018document} and the enumeration problem associated to it~\cite{FlorenzanoRUVV20}.
Recently, an extension of document spanners has been proposed to enhance the extraction process with annotations~\cite{DoleschalKMP20}.
These annotations serve as auxiliary information of the extracted data such as confidence, support, or confidentiality measures. 
To extend spanners, this framework follows the approach of provenance semiring by annotating the output with elements from a semiring and propagate the annotations by using the semiring operators. 
Next we give the core definitions of~\cite{DoleschalKMP20} to state then the implications of our main results. 

We start by defining the central elements of document spanners: documents and spans. 
Fix a finite alphabet $\Sigma$. A document over $\Sigma$ (or just a document) is a string $d = a_1 \ldots a_n \in \Sigma^*$ and a span is pair $s = \mspan{i,j}$ with $1 \leq i \leq j \leq n+1$.
A span represents a continuous region of
$d$, whose content is the substring of $d$ from positions $i$ to
$j-1$. Formally, the content of span $\mspan{i,j}$ is defined as $d\mspan{i,j} = a_i \ldots a_{j-1}$; if $i=j$, then $d\mspan{i,i} = \epsilon$.
Fix a finite set of variables  $\xset$.
A mapping $\mu$ over $d$ is a function from $\xset$ to the spans of $d$.  
A document spanner (or just spanner) is a function that transforms each document $d$ into a set of mappings over $d$. 

To annotate mappings, we need to introduce semirings. A semiring $(K, \ksum, \kprod, \kzero, \kone)$ is an algebraic structure where $K$ is a non-empty set, $\ksum$ and $\kprod$ are binary operations over $K$, and $\kzero, \kone \in K$. Furthermore,  $\ksum$ and $\kprod$ are associative, $\kzero$ and $\kone$ are the identities of $\ksum$ and $\kprod$ respectively, $\ksum$ is commutative, $\kprod$ distributes over $\ksum$, and $\kzero$ annihilates $K$ (i.e. $\forall k \in K. \, \kzero \kprod k = k \kprod \kzero = \kzero$). We will use $\bigksum_X$ or $\bigkprod_X$ for the $\ksum$- or $\kprod$-operation over all elements in some set $X$, respectively.
An ordered semiring $(K, \ksum, \kprod, \kzero, \kone, \preceq)$ is a semiring extended with a total order $\preceq$ over $K$ such that $\preceq$ preserves $\ksum$ and $\kprod$, namely, $k_1 \preceq k_2$ implies $k_1 * k \preceq k_2 * k$ for $* \in \{\ksum, \kprod\}$. From now on, we will assume that all semirings are ordered. 
A semifield~\cite{golan2013semirings} is a semiring $(K, \ksum, \kprod, \kzero, \kone)$ where each $k \in K \setminus \{\kzero\}$ has a multiplicative inverse, i.e. $(K \setminus \{\kzero\}, \kprod, \kone)$ form a group.
Examples of ordered semifields are the tropical semiring $(\bbZ\cup \{\infty\}, \min, +, \infty , 0, \leq)$ and the semiring of non-negative rational numbers $(\mathbb{Q}_{\geq 0}, +, \times, 0, 1, \leq)$.

Fix a semiring $(K, \ksum, \kprod, \kzero, \kone)$. Let $\xset$ be a set of variables and define $\mathcal{C}(\xset) = \{\Open{x}, \Close{x} \mid x \in \xset\}$. To define spanners with annotations, we use the formalism of weighted variable set automata~\cite{DoleschalKMP20} which defines the class of all regular spanners with annotations, also called regular annotators.
A weighted variable set automaton (wVA) over $K$ is a tuple $\cA = (\xset, Q, \delta,  I, F)$ such that $\xset$ is a finite set of variables, $Q$ is a finite set of states, $\delta: Q \times (\Sigma \cup \mathcal{C}(\xset)) \times Q \rightarrow K$ is a weighted transition function and $I: Q \rightarrow K$ and $F:Q \rightarrow K$ are the initial and final weight functions, respectively.
A run $\rho$ over a document $d = a_1  \,\cdots\, a_n$ is a sequence of the form:
\[
\rho := (q_0, i_0) \ \trans{o_1} \ (q_1, i_1) \ \trans{o_2} \ \ldots \ \ \trans{o_m} \ (q_m, i_m)
\]
where (1) $1 = i_0 \leq i_1 \leq \cdots \leq i_m = n+1$, (2) each $q_j \in Q$ with $I(q_0) \neq \kzero \neq F(q_m)$, (3) $\delta(q_j, o_{j+1}, q_{j+1}) \neq \kzero$, and (4) $i_{j+1} = i_{j}$ if $o_{j+1} \in \mathcal{C}(\xset)$ and $i_{j+1} = i_j +1$ otherwise. 
In addition, we say that a run $\rho$ is valid if for every $x \in \xset$ there exists exactly one index $i$ with $o_i = \Open{x}$, exactly one index $j$ with $o_j = \Close{x}$, and $i < j$.
We denote by $\run_{\cA}(d)$ the set of all valid runs of $\cA$ over $d$. 
Note that for some wVA $\cA$ and document $d$ there could exist runs of $\cA$ over~$d$ that are not valid. We say that $\cA$ is functional if every run $\rho$ of $\cA$ over $d$ is valid for every document $d$. Given that decision problems associated to non-functional variable-set automata have been shown to be \textsc{NP}-hard~\cite{freydenberger2018joining, maturana2018document}, from now on we assume that all wVA are functional.

A valid run $\rho$ like above naturally defines a mapping $\mu^{\rho}$ over $\xset$ that maps each $x$ to the span $\mspan{i_{j},i_{j'}}$ where $o_{i_j} = \Open{x}$ and $o_{i_{j'}} = \Close{x}$.
Furthermore, we can associate a weight in $K$ to $\rho$ by multiplying all the weights of the transitions, formally,
\[
W(\rho) := I(q_0) \kprod \bigkprod_{j=1}^m \delta(q_j, o_{j+1}, q_{j+1}) \kprod F(q_m)
\]
We define the set of output mappings of $\cA$ over $d$ as $\sem{\cA}(d) = \{\mu^\rho \mid \rho \in \run_{\cA}(d)\}$.
Given a mapping $\mu \in \sem{\cA}(d)$ we associate a weight $W_{\cA,d}(\mu) = \bigksum_{\rho \in \run_{\cA}(d): \mu = \mu^{\rho}} W(\rho)$. 
Intuitively, each $\mu \in \sem{\cA}(d)$ contains relevant data extracted by $\cA$ from $d$ and $W_{\cA,d}(\mu)$ is the additional information attached to $\mu$ obtained by $\cA$ from $d$ in the extraction process, e.g. confidence or support. 

In~\cite{DoleschalKMP20}, the problem of ranked annotator enumeration was proposed, which for the sake of completeness we present next\footnote{In~\cite{DoleschalKMP20} they considered positively ordered semiring, which is slightly more general that the notion of ordered semiring used here.}:
\vspace{.1cm}
\begin{center}
	\framebox{
	\begin{tabular}{rl}
		\textbf{Problem:} & $\RAENUM$\\
		\textbf{Input:} & A wVA $\cA$ over an ordered semiring $K$ \\
		& and a document $d$.  \\
		\textbf{Output:} & Enumerate all $\mu_1, \ldots, \mu_k \in \sem{\cA}(d)$ \\
		& without repetitions and such that \\
		&  $W_{\cA,d}(\mu_1) \preceq W_{\cA,d}(\mu_{i+1})$.
	\end{tabular}
}
\end{center}
\vspace{.1cm}
$\RAENUM$ was studied in~\cite{DoleschalKMP20} and an enumeration algorithm was provided with polynomial preprocessing and polynomial delay in terms of $|\cA|$ and $|d|$. By using the framework of MSO cost functions, we can give a better algorithm for a special case of $\RAENUM$. 
We say that a wVA $\cA$ is unambiguous if, for every document $d$ and $\mu \in \sem{\cA}(d)$, there exists at most one run $\rho \in \run_{\cA}(d)$ such that $\mu = \mu^{\rho}$. Of course, the similarity between cost transducers and wVA is more or less clear, although the former works over groups and wVA works over semirings. For this reason, we restrict wVA to semifields and give the following result.
\begin{corollary}
	The problem $\RAENUM$ can be solved with $|\cA|\cdot|d|$ preprocessing time and $\log(|\cA| \cdot |d|)$-delay when $\cA$ is unambiguos and $K$ is an ordered semifield. 
\end{corollary}
Although the previous result is a restricted case of $\RAENUM$ and a direct consequence of Theorem~\ref{theo:renumt}, to the best of our knowledge this is the first non-trivial ranked enumeration algorithm proposed for the framework of document spanner.

\subsection{Complex event processing}


\begin{figure*}
	\centering
	\begin{subfigure}{.6\textwidth}
		\centering
		{
			\vspace{3mm}
			\resizebox{\textwidth}{!}{
			\begin{tabular}{|c|c|c|c|c|c|c|c|c|c|c}\hline
				type  &$H$&$T$&$T$&$H$&$H$&$T$&$T$&$T$&$H$ & \ldots \\ \hline
				value & 25 & 25& 20& 25& 40& 42& 25& 70& 18 & \ldots \\ \hline
				index & 1 & 2 & 3 & 4 & 5 & 6 & 7 & 8 & 9 & \ldots \\ \hline
			\end{tabular}}
			\vspace{2mm}}
	\end{subfigure}%
	\begin{subfigure}{.4\textwidth}
		\centering
		\resizebox{\textwidth}{!}{
		\begin{tikzpicture}[->,>=stealth, semithick, auto, initial text= {}, initial distance= {3mm}, accepting distance= {4mm}, node distance=2.2cm, semithick]
		\tikzstyle{every state}=[draw=black,text=black,inner sep=0pt, minimum size=8mm]
		\node[initial,state]	(1) 				{$q_1$};
		\node[state]			(2) [right of=1]	{$q_2$};
		\node[accepting,state]	(3) [right of=2]	{$q_3$};
		\path
		(1)
		edge 				node {$P \mid \amark$} (2)
		edge [loop above] 	node {$\texttt{TRUE} \mid \umark$} (1)
		(2)
		edge [in=100,out=130,loop] node[pos=0.55] {$P' \mid \amark$} (2)
		edge [in=50,out=80,loop] node[pos=0.45] {$\texttt{TRUE} \mid \umark$} (2)
		edge node {$P \mid \amark$} (3);
		\end{tikzpicture}}
	\end{subfigure}
	\caption{At the left, a stream $S$ of events measuring temperature and humidity. ``value'' contains degrees and humidity for $T$- and $H$- events, respectively. At the right, a complex event automaton where $P := \type[H]$ and $P':= \type[T] \wedge value > 40$.}
	\label{fig:cep}
\end{figure*}
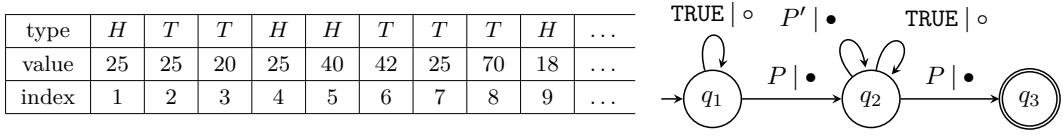

In the dynamic query evaluation~\cite{BerkholzKS17,IdrisUVVL20} setting, the goal is to maintain the result of a query under updates to the underlying database.
The most used approach is to maintain a data structure that represents the partial evaluation of the query.
Then for each new update, the data structure is minimally updated (e.g., constant time), and the fresh outputs are enumerated with some strong guarantees in the delay (e.g. constant delay).
Complex event processing (CEP) is an orthogonal approach to dynamic query evaluation~\cite{UgarteV18}, in which the updates are represented by an infinite stream of data items, called events, and where the time, represented by the order of the events in the stream, plays a significant role.
For these reasons, in CEP the stream resembles more an infinite string and the operators of the query language are closely connected to regular expressions.
Moreover, similar techniques are used for its evaluation, in particular the use of intermediate automata models, and the techniques developed for constant delay enumeration of MSO naturally apply in this context. We continue this subsection by introducing the main definitions of CEP, to then extend the (offline) ranked enumeration setting to an online version. Towards the end, we give the main results of our ranked enumeration algorithm in this context.


Let $\aset$ be an infinite set of attribute names and $\dset$ an infinite set of values. 
A database schema $\cR$ is a finite set of relation names, where each relation name $R \in \cR$ is associated to a tuple of attributes denoted by $\att(R)$. If $R$ is a relation name, then an $R$-tuple is a function $t:\att(R) \rightarrow \dset$ and define its type as $\type(t) = R$. For any relation name~$R$, $\tuples(R)$ denotes the set of all possible $R$-tuples. Similarly, $\tuples(\cR)=\bigcup_{R \in \cR}\tuples(R)$ for any database schema~$\cR$. 
Given $\cR$, a unary predicate is a set $P \subseteq \tuples(\cR)$.
We denote by $\upset$ a finite and fix set of unary predicates over $\cR$. We also assume that $\upset$ contains a predicate $\type[R] = \{t \mid \type(t) = R\}$ for every $R \in \cR$, and $\upset$ is closed under conjunction (i.e., $P \cap P' \in \upset$ for every $P, P' \in \upset$).
Furthermore, we assume that, for every tuple $t$ and $P \in \upset$, verifying if $t \in P$ takes constant time in the RAM computational model\footnote{In other words, we assume that the time cost of verifying if a tuple satisfies a predicate is not relevant for the time cost of the algorithm.}.  

Fix a schema $\cR$. A stream $S$ is just an infinite sequence $S = t_1 t_2 \ldots$ where $t_i \in \tuples(\cR)$. Given a position $n$, we write $S_n$ to denote the prefix $t_1 \ldots t_n$ of $S$.
In CEP, a tuple $t_i$ of $S$ is called an \emph{event} and a finite subset $\{t_{i_1}, \ldots, t_{i_k}\}$ is called a \emph{complex event} of $S$. 
For the sake of simplification, in this paper we define a complex event as non-empty finite set of positions $C \subseteq \mathbb{N}\setminus \{0\}$. This set $C$ naturally defines the set of events as $\{t_i \mid i \in C\}$. 
Finally, we write $\min(C)$ and $\max(C)$ for the first and last position in $C$, respectively. 

To define complex events over streams, we use the model of complex event automata~\cite{grez2019} which is expressible enough to define all (regular) complex event queries over streams~\cite{grez2019}.
Fix a schema $\cR$ and a set of unary predicates $\upset$ (defined as above).
A \emph{complex event automaton} (CEA) is a tuple $\cA = (Q, \Delta, I, F)$ where $Q$ is a finite set of states, $\Delta \subseteq Q \times \upset \times \{\umark, \amark\} \times Q$ is a finite transition relation, and $I, F \subseteq Q$ are the set of initial and final states, respectively. Intuitively, the elements $\{\umark, \amark\}$ indicate whether or not the element used to take the transition will be part of the output. 
Given a stream $S = t_1 t_2 \ldots$, a run $\rho$
of $\cA$ over $S$ of length $n$ is a sequence of transitions
$\rho: q_0 \ \trans{P_1 / m_1} \ q_1 \  \trans{P_2 / m_2} \ \cdots \ \trans{P_n / m_n} \ q_{n}$
such that $q_0 \in I$, $t_i \in P_i$ and $(q_{i-1}, P_{i}, m_i, q_{i}) \in \Delta$ for every $i \in [1,n]$.
We say that $\rho$ is \emph{accepting} if $q_{n} \in F$. We write $\run_n(\cA, S)$ to denote the set of all accepting runs of $\cA$ over $S$ of length $n$.
Further, we define the complex event induced by $\rho$ as $C_\rho = \{i \in [1, n] \mid m_i = \amark\}$.
Given a stream $S$ and $n > 0$, we define the set of complex events of $\cA$ over $S$ at position $n$ as $\sem{\cA}_n(S) = \{C_\rho \mid \rho \in \run_n(\cA, S) \}$.
Finally, we say that $\cA$ is unambiguous if for every stream $S$, position $n$, and $C \in \sem{\cA}_n(S)$, there exists exactly one run $\rho$ of $\cA$ over $S$ of length $n$ such that $C = C_\rho$.

\begin{example}\label{ex:cea}
Consider the following example from~\cite{grez2019} about a network of sensors measuring the temperature and humidity in a farm.  The stream of data is composed by two kinds of events: tuples of type $T$ or $H$ whose attributes ``value'' contain temperature or humidity values, respectively, measured by a sensor. The left side of Figure~\ref{fig:cep} shows an example of a stream $S$ generated by this network, where each column represents an event with its type, value, and index in the stream.
The right side of Figure~\ref{fig:cep} shows an example of a CEA $\cA_0$ over the schema $\cR = \{T,H\}$. This CEA captures complex events whose first and last events are of type $H$ and the events in-between are of type $T$ with value greater than $40$. Note that the transitions labeled with $\umark$ have $\texttt{TRUE}$ as predicate, meaning that $\cA_0$ can decide to skip events arbitrarily. If we run $\cA_0$ over~$S$, we can check that the complex events captured at position $9$ are $\sem{\cA}_9(S) = \{ \{5, 6,8,9\}, \{4,6,8,9\}, \{1,6,8,9\}, \{5,6,9\}, \ldots \}$.
\end{example}

Given a CEA $\cA$ and a stream $S$, the evaluation problem consists in enumerating $\sem{\cA}_n(S)$ at each position $n$. Of course, ranked enumeration can also be applied in this setting. More specifically, for any stream $S = t_1 t_2 \ldots$ and for each position $n$ we can define a structure $S_n = ([n], \leq, (P_S)_{P \in \upset})$ such that $P_S(i)$ is true if, and only if, $t_i \in P$. To define the cost of each complex event at position $n$, we can use a weighted MSO formula $\alpha(X)$ over $\upset$ such that the cost of $C \in \sem{\cA}_n(S)$ is equal to $\sem{\alpha}(S_n, \sigma[C])$ where $\sigma[C] = \sigma[X \mapsto C]$. 
Then, given a weighted MSO formula $\alpha(X)$ over $\upset$ we can state the ranked enumeration problem of CEA as follows.
\vspace{.3cm}
\begin{center}
	\framebox{
		\begin{tabular}{rl}
			\textbf{Problem:} & $\RENUMCEA[\alpha]$\\
			\textbf{Input:} & A CEA $\cA$ and a stream $S = t_1 t_2 \ldots $.  \\
			\textbf{Output:} & After reading $t_1 \ldots t_n$, enumerate all \\ & $C_1, \ldots, C_k \in \sem{\cA}_n(S)$ without repetitions  and \\ & such that $\sem{\alpha}(S_n, \sigma[C_i]) \preceq \sem{\alpha}(S_n, \sigma[C_{i+1}])$. 
		\end{tabular}
	}
\end{center}
\vspace{.3cm}
\begin{example}\label{ex:cefunction}
A complex event $C$ can be placed arbitrarily far from the current position $n$ in the stream, which might give $C$ less importance.
Hence, a useful cost function is to measure the distance between the first event in $C$ and the current position $n$. We can define this distance with a weighted MSO formula over $\bbZ$ as follows:
\[
\alpha_3 \ := \ \gsum z.  [(\exists x. x \in X \wedge x \leq z) \mapsto 1]
\]
One can check that $\sem{\alpha_3}(S_n, \sigma[X \mapsto C]) = n-\min(C)+1$ for every stream $S$, position $n$, and complex event $C$.
Thus, if we enumerate $\sem{\cA}_n(S)$ ranked by $\alpha_3$, the complex events that are ``closer'' to $n$ in the stream will be enumerated first. 
\end{example}

Although weighted MSO formulas and ranked enumeration can be naturally adjusted in this context, the guarantees of efficiency of the enumeration process is a bit more subtle. Intuitively, each time that a new event $t_n$ arrives, we do not want to preprocess the whole prefix $S_{n-1}$ again to then start the enumeration phase. Given that we already processed $S_{n-1}$, we would like to keep a data structure $D$ such that, whenever the new event $t_n$ arrives, we update $D$ with $t_n$ taking time proportional to $|t_n|$, and then enumerate the new outputs that have been found. More precisely, assume that the stream $S$ is read by calling a special instruction ${\tt yield}_S$ that returns the next element of $S$ each time it is called. Then, we say that $\mathcal{E}$ is a \emph{streaming enumeration algorithm} for a CEA $\cA$ over a stream $S$ if $\mathcal{E}$ maintains a data structure $D$ in memory such that:
\begin{enumerate} 
	\item \label{item:cda1} between any two calls to ${\tt yield}_S$, the data structure $D$ is updated with $t$ where $t$ is the tuple returned by the first of such calls, and
	\item \label{item:cda2} after calling ${\tt yield}_S$ $n$ times, the set $\sem{\cA}_n(S)$ can be enumerated from $D$.
\end{enumerate}
We assume here that the enumeration phase at position $n$ is exactly the same as the enumeration phase of a (normal) enumeration algorithm (see Section~\ref{sec:preliminaries}). Furthermore, we say that $\mathcal{E}$ has delay $g: \mathbb{N}^2 \rightarrow \mathbb{N}$ when there exists a constant $F$ such that the delay $\delay_i(\cA, S_n)$ between two complex events $C_i$ and $C_{i+1}$ in $\sem{\cA}_n(S)$ is bounded by $F \times |C_{i+1}| \times g(|\cA|, |S_n|)$. 
We say that $\mathcal{E}$ has update-time $f: \mathbb{N}^2 \rightarrow \mathbb{N}$ if there exists a constant $G$ such that the number of instructions that $\mathcal{E}$ executes during the update of $D$ with $t$ is bounded by $G \times f(|\cA|, |t|)$. In particular, the update of $D$ does not depend on the number of events seen so far (i.e., does not depend on $S_n$).
We say that $\RENUMCEA[\alpha]$ can be solved with update-time $f$ and delay $g$ if there exists a streaming enumeration algorithm $\mathcal{E}$ with update-time $f$ and delay $g$, that enumerates $\sem{\cA}_n(S)$ in increasing order according to $\alpha$ at each position $n$. 

It is important to notice that the existence of an enumeration algorithm for the ranked enumeration problem of MSO does not imply the existence of a streaming enumeration algorithm. 
However, shown in Section~\ref{sec:renumalgo}, our ranked enumeration algorithm maintains a data structure to process the input in such a way that it is updated ``one letter at a time'' and, at any moment of the preprocessing phase, all outputs until that moment can be efficiently enumerated. That is, we can derive the following result:

\begin{theorem}\label{theo:renumcea}
	The problem $\RENUMCEA[\alpha]$ can be solved with $2^{|\cA|}\cdot |t|$ update-time and $\log(|\cA| \cdot |S_n|)$-delay. Furthermore, if $\cA$ is unambiguous, then $\RENUMCEA[\alpha]$ can be solved with $|\cA|\cdot |t|$ update-time and $\log(|\cA| \cdot |S_n|)$-delay. 
\end{theorem}


For this result, the preprocessing part of Algorithm~\ref{alg:evaluation} has to be modified to call the enumeration procedure after reading each event in the stream, instead of only at the end.

An important property of streams in CEP is that the relevance of events and complex events rapidly decays over time. To illustrate this, recall the setting of Example~\ref{ex:cea}. Arguably, a user would prefer the complex event $C_1=\{5,6,8,9\}$ over $C_2=\{1,6,8,9\}$ because its time interval is shorter and closer to the current time $9$. For this reason, CEP query languages usually include time operators that filter out the complex events that are not inside a sliding window~\cite{cugola2012processing}. Formally, given a CEA $\cA$ and a number $T$ (encoded in binary) consider the query $Q := \cA \WITHIN T$ such that, for every stream $S$ and position $n$, $Q$ defines the set of complex events $\sem{\cA \WITHIN T}_n(S) = \{C \in \sem{\cA}_n(S) \mid n - \min(C) \leq T\}$. In other words, it considers only those $C$ captured by $\cA$ that are contained within the last $T$ events of the stream.
Interestingly, we can use Theorem~\ref{theo:renumcea} to efficiently evaluate queries of the form $\cA \WITHIN T$. Indeed, if we evaluate $\cA$ over a stream $S$, rank each complex event with the cost function $\alpha_3$ of Example~\ref{ex:cefunction}, and enumerate all complex events in increasing order up to cost $T$, we will have enumerated all complex events in $\sem{\cA \WITHIN T}_n(S)$. Thus, we easily get the following corollary for the evaluation of CEP queries over sliding windows. 

\begin{corollary}
	For each CEA $\cA$ and value $T$ (in binary), $\cA \WITHIN T$ can be evaluated with $2^{|\cA|}\cdot |t|$ update-time and $\log(|\cA| \cdot |S_n|)$-delay. Furthermore, if $\cA$ is restricted to be unambiguous, then it can be evaluated with $|\cA|\cdot |t|$ update-time and $\log(|\cA| \cdot |S_n|)$-delay. 
\end{corollary}

It is important to clarify two facts about the previous result. First, the result can be extended to CEP queries over sliding windows when time is continuous (e.g. in seconds) by slightly modifying our evaluation algorithm. 
Second, the advantage of the previous result is that the evaluation process does not depend on the length $T$ of the sliding window. Although the length $T$ reduces the number of events that a query needs to ``see'' for the evaluation, given that $T$ is in binary (or time is continuous), the sliding window could contain a huge number of events during the evaluation process.

\section{Proof of Proposition~\ref{prop:WMSO-CT}} \label{apx:proofOfWMSO-CT}

\newcommand{\lfirst}{\operatorname{first}}
\newcommand{\llast}{\operatorname{last}}
\newcommand{\lsucc}{\operatorname{succ}}
\newcommand{\lrun}{\operatorname{run}}
\newcommand{\linit}{\operatorname{init}}
\newcommand{\ltrans}{\operatorname{trans}}
\newcommand{\lfinal}{\operatorname{final}}

The following is a proof taken from Theorems~4.1 and 5.3 of \cite{KreutzerR13} and adapted to our setting.
Instead of citing the results there, we decided to add the complete proof to keep the paper self-contained.
First, lets fix an ordered group $(\bbG, \gop, \gid, \preceq)$.

$(\Rightarrow)$ Consider a cost transducer $\cT = (Q,\Delta,\kappa,I,F)$ and let $V$ be the output alphabet.
We define formulas $\varphi_\cT$ and $\alpha_\cT$ with free variables $V$, where the former encodes a run of $\cT$ and the latter calculates the cost of such run.
In the following, let $\Delta_I$ be the set of transitions coming from an initial state and $\Delta_F$ be the set of transitions going to a final state.
Moreover, for every $U \in V$, define the set $\Delta_U = \{(p,a,V',q) \in \Delta \mid U \in V'\}$ of transitions that output $U$.

We begin by defining $\varphi_\cT$.
First of all, we introduce some auxiliary formulas: $\lfirst(x) := \forall y.\, x \leq y$; $\llast(x) := \forall y.\, y \leq x$; $\lsucc(x,y) := x \leq y \land y \nleq x \land \forall z. (z \leq x \lor y \leq z)$.
These denote the first element, last element and successor relation, respectively, according to the order $\leq$ of the word structure.
We encode a run $\rho$ by defining, for each $t = (p,a,V,q)\in \Delta$, a variable $X_t$ such that $i \in X_t$ if the $i$-th transition of $\rho$ is $t$.
Let $\bar{X} = \{X_t \mid t \in \Delta\}$.
Now, we define the predicate $\lrun(\bar{X},V)$ that is satisfied only when $\bar{X}$ defines an accepting run in $\cT$ such that its output defines an encoding equal to the assignment of $V$.
\[
\begin{split}
\lrun(\bar{X},V) \ := \ \ 
& \forall x. \bigvee_{t \in \Delta} (x \in X_t \wedge \bigwedge_{t' \neq t} x \notin X_{t'})  \ \wedge\\
& \bigwedge_{(p,a,U,q)\in \Delta} \forall x. (x \in X_{(p,a,U,q)} \ \rightarrow \ x \in P_a) \ \land \\
& \forall x. \forall y. \Big( \lsucc(x,y) \ \rightarrow \ \bigvee_{(p,a,U,q), (q,b,W,r) \in \Delta} x \in X_{(p,a,U,q)} \ \wedge \ y \in X_{(q,b,W,r)} \Big) \ \wedge \\
& \exists x. \Big( \lfirst(x) \ \wedge \ \bigvee_{t \in T_I} x \in X_t \Big) \ \wedge \ \exists x. \Big(\llast(x) \ \wedge \ \bigvee_{t \in T_F} x \in X_t \Big) \\
& \bigwedge_{U \in V} \Big( \forall x.  x \in U \leftrightarrow \bigvee_{t \in \Delta_U} x \in X_t \Big)
\end{split}
\]
The first line  makes sure that exactly one transition is used per position; the second makes sure that each transition is taken only if the input letter is correct; the third checks that the transitions form a path; the fourth checks the initial and final states and the last checks that the assignment output by the run corresponds to the assignment of $V$.

Then, we define our first formula:
\[
\varphi_\cT(V) := \exists \bar{X}. \lrun(\bar{X},V)
\]
Note that $\varphi_\cT$ guesses a run and, because $\cT$ is unambiguous, that run is unique for every assignment of $V$.
Then it holds that, for every word $w$, $\sem{\varphi}(w) = \sem{\cT}(w)$. 

Now, to define $\alpha_\cT$, we define cost formulas that capture the costs of the initial state, the transitions and the final states:
\[
\begin{split}
	\linit(V) \ :=& \ \biggop_{(p,a,U,q)\in T_I} \Big( \exists \bar{X}. \big( \lrun(\bar{X},V) \wedge \exists x. (\lfirst(x) \wedge x \in X_{(p,a,U,q)}) \big) \mapsto I(p) \Big) \\
	\ltrans(x,V) \ :=& \ \biggop_{(p,a,U,q)\in T} \Big( \exists \bar{X}. \big( \lrun(\bar{X},V) \wedge x \in X_{(p,a,U,q)} \big) \mapsto \kappa((p,a,U,q)) \Big) \\
	\lfinal(V) \ :=& \ \biggop_{(p,a,U,q)\in T_F} \Big( \exists \bar{X}. \big( \lrun(\bar{X},V) \wedge \exists x. (\llast(x) \wedge x \in X_{(p,a,U,q)}) \big) \mapsto F(q) \Big)
\end{split}
\]
The $trans$ formula is supposed to receive a position $x$ of the domain and retrieve the cost of the $x$-th transition of the guessed run.
Then, we define our second formula:
\[
\alpha_\cT(V) \ := \ \linit(V) \ \gop \ \gsum x. \, \ltrans(x,V) \ \gop \ \lfinal(V)
\]
It is then easy to check that $\costt{\cT}(w,\sigma) = \sem{\alpha_\cT}(w,\sigma)$ for every word $w$ and $V$-assignment $\sigma$.

$(\Leftarrow)$ Consider an MSO formula $\varphi$ and a weighted MSO formula $\alpha$.
Given an alphabet $\Sigma$ and a set of (second and first order) variables $V$, let $\Sigma_V = \Sigma \times 2^{V}$ be the alphabet used to encode words and assignments.
Moreover, by abuse of notation, let $(w,\sigma) \in \Sigma_V$ be the word encoding $w$ and $\sigma$, i.e., $(w,\sigma)=(a_1,V_1)(a_2,V_2)\ldots (a_n,V_n)$, where $a_1\ldots a_n = w$ and $X \in V_i$ iff $i \in \sigma(X)$.
The following Lemma is a well-known result that relates the expressibility of MSO and DFA.
\begin{lemma} \label{MSO-DFA}
	For every MSO formula  $\psi$ over $\Sigma$ with free variables $V$ there exists a DFA $\cA_\psi$ over $\Sigma_V$ such that for every word $w \in \Sigma^*$ and assignment $\sigma$ we have:
	\[
	(w,\sigma) \vDash \psi(V) \ \text{ iff } \ (w,\sigma) \in \cL(\cA_\psi)
	\]
\end{lemma}
where $\cL(\cA)$ is the set of words accepted by the DFA $\cA$.
Therefore, regarding the MSO formula $\varphi$, because of Lemma~\ref{MSO-DFA} we know there exists a DFA $\cA_\varphi$ such that $ (w,\sigma) \vDash \varphi(V)$ iff $(w,\sigma) \in \cL(\cA_\varphi)$.
Next, we build a cost transducer $\cT_\alpha$ for $\alpha$, then make the cross-product of it with $\cA_\varphi$ and obtain the final cost transducer.

In particular, the cost transducer $\cT_\alpha = (Q_\alpha,\Delta_\alpha,\kappa_\alpha,I_\alpha,F_\alpha)$ is equivalent to $\alpha$ in the sense that it satisfies $\sem{\alpha}(w,\sigma) = \costt{\cT_\alpha}(w,\sigma)$ for every word $w$ and assignment $\sigma$.
Moreover, if al subformulas of $\alpha$ are of the form $\varphi \mapsto g$ or $\alpha_1 \gop \alpha_2$, then $\cT_\alpha$ is deterministic, meaning that it has a single accepting state $q_0 \in \dom(I_\alpha)$ and, for every $p\in Q_\alpha$, $a \in \Sigma$ and $V' \subseteq V$ there is at most one transition $(p,a,V',q) \in \Delta_\alpha$.
Usually, we use notation $\cT = (Q,\delta,\kappa,I,F)$ to denote that $\cT$ is deterministic.
Otherwise, if $\alpha$ also contains operation $\gsum. x$, then $\cT_\alpha$ is unambiguous.

We build $\cT_\alpha$ by induction over the structure of $\alpha$.
The base case is when $\alpha = [\varphi' \mapsto g]$, with $\varphi'$ being an MSO formula and $g \in \bbG$.
Consider the DFA $\cA_{\varphi'} = (Q',\delta',q_0',F')$ given by Lemma~\ref{MSO-DFA}.
Then, we build $\cT_\alpha = (Q',\delta_\alpha,\kappa_\alpha,I_\alpha,F_\alpha)$ such that:
\begin{itemize}
	\item $\delta_\alpha = \{(p,a,V',q) \mid (p,(a,V'),q) \in \delta'\}$,
	\item $\kappa_\alpha(t) = \gid$ for every $t \in \delta_\alpha$,
	\item $I_\alpha(q_0') = \gid$, and
	\item $F_\alpha(q) = g$ for every $q \in F'$.
\end{itemize}
It is not hard to see that $\cT$ is deterministic and that for every word $w$ and assignment $\sigma$, $\sem{\alpha}(w,\sigma) = \costt{\cT_\alpha}(w,\sigma)$.

Now, we consider the case where $\alpha = \alpha_1 \gop \alpha_2$.
By induction, consider that there are cost transducers $\cT_i = (Q_i,\Delta_i,\kappa_i,I_i,F_i)$ equivalent to $\alpha_i$, for $i \in \{1,2\}$.
Then, we build $\cT_\alpha = (Q_\alpha,\Delta_\alpha,\kappa_\alpha,I_\alpha,F_\alpha)$ as the cross product of $\cT_1$ and $\cT_2$ in the following way:
\begin{itemize}
	\item $Q_\alpha = Q_1 \times Q_2$,
	\item $\Delta_\alpha = \{((p_1,p_2),a,V',(q_1,q_2)) \mid (p_1,a,V',q_1) \in \Delta_1 \land (p_2,a,V',q_2) \in \Delta_2\}$,
	\item $\kappa_\alpha(((p_1,p_2),a,V',(q_1,q_2))) = \kappa_1(p_1,a,V',q_1) \gop \kappa_2(p_2,a,V',q_2)$,
	\item $I_\alpha((q_1,q_2)) = \gid$ for all $q_1 \in \dom(I_1)$ and $q_2 \in \dom(I_2)$, and
	\item $F_\alpha((q_1,q_2)) = F_1(q_1) \gop F_2(q_2)$ for all $q_1 \in \dom(F_1)$ and $q_2 \in \dom(F_2)$.
\end{itemize}
Again, for every word $w$ and assignment $\sigma$, $\sem{\alpha}(w,\sigma) = \costt{\cT_\alpha}(w,\sigma)$.
Moreover, if $\cT_1$ and $\cT_2$ are deterministic (unambiguous), then $\cT_\alpha$ is deterministic (unambiguous, respectively).

Finally, we consider the case where $\alpha = \gsum x. \, \alpha'$.
Let $\cT' = (Q',\delta',\kappa',I',F')$ be the cost transducer equivalent to $\alpha'$.
Because there are no nested $\gsum x$ operators, we know that $\cT'$ is deterministic.
Moreover, because of the construction, for every run $\rho$ ending in some state $q_f \in \dom(F')$, all costs are added by the final state, i.e. $\kappa'(t) = \gid$ for all $t \in \delta'$, $I(q) = \gid$ for all $q \in \dom(I)$, and $\kappa(\rho) = F(q_f)$.
We make use of this property in the following construction.
For simplicity, assume that $V = \{x\}$, so that transitions of $\cT'$ either have the form $(p,a,\{x\},q)$ or $(p,a,\emptyset,q)$; we call the latter $\emptyset$-transitions.
Next, we define $\cT_\alpha = (Q_\alpha,\Delta_\alpha,\kappa_\alpha,I_\alpha,F_\alpha)$ equivalent to $\alpha$:
\begin{itemize}
	\item $Q_\alpha = Q' \times 2^{Q'\times Q'}$,
	\item $I_\alpha(q,\emptyset) = \gid$ for all $q \in \dom(I')$
	\item $F_\alpha(q,R) = \gid$ if $R$ has the form $R = \{(q_1,q_1),\ldots,(q_k,q_k)\}$,
	\item $t = ((q_1,R_1),a,\emptyset,(q_2,R_2)) \in \Delta_\alpha$ and $\kappa(t) = s$ if there exists $q_f \in Q'$ such that
	\begin{itemize}
		\item $F'(q_f) = s$
		\item $(q_1,a,\emptyset,q_2) \in \delta'$, and
		\item $R_2 = \{(p_2,q_f') \in Q^2 \mid \exists (p_1,q_f') \in R_1. (p_1,a,\emptyset,p_2)\in\delta'\} \cup \{(q,q_f) \mid (q_1,a,\{x\},q) \in \delta'\}$.
	\end{itemize}
\end{itemize}
The motivation behind this construction is that, for each $(q,R) \in Q_\alpha$, $q$ keeps the state of the (only) run of $\cT'$ considering that $x$ has not been assigned, while $R$ keeps pairs $(p,q_f)$ such that $p$ follows a run that has already assigned $x$ and $q_f$ is the ``guessed'' final state of such run.
Moreover, every time a final state is guessed, its value $F(q_f)$ is added to the cost of the run.
The guess then is verified by checking that each final state is reached as expected.

To check that $\cT_\alpha$ is unambiguous, consider by contradiction that there are two accepting runs
\[
\rho_i = (q_0,\emptyset) \trans{a_1/\emptyset} (q_1,T_1^i) \trans{a_2/\emptyset} \cdots \trans{a_n/\emptyset}(q_n,T_n^i)
\]
for $i \in \{1,2\}$.
Then, let $j \leq n$ be the smallest position at which both runs differ, namely $T_j^1 \neq T_j^2$ and $T_{j-1} = T_{j-1}^1 = T_{j-1}^2$.
By looking at the definition of $T_j^1$ and $T_j^2$, we can see that both have the same set in the first part, defined completely by $T_{j-1}$, chich is $S = \{(p_2,q_f') \in Q^2 \mid \exists (p_1,q_f') \in T_{j-1}. (p_1,a,\emptyset,p_2)\in\delta'\}$).
For the second part, let $q_f^1$ and $q_f^2$ be the final states guessed by the transitions in $\rho_1$ and $\rho_2$.
Then, for $i \in \{0,1\}$, $T_j^i = S \cup \{(q,q_f^i) \mid (q_{j-1},a_j,\{x\},q) \in \delta'\}$.
Because $\cT'$ is deterministic, the only way for the sets to differ is if $q_f^1 \neq q_f^2$.
Now, let $p \in \cT'$ be the state reached from $q$ after reading the remaining suffix of $w$, i.e. $w[j+1]\ldots w[n]$, using only $\emptyset$-transitions.
From the construction, $(p,q_f^1)$ must be in $T_n^1$ and $(p,q_f^2)$ must be in $T_n^2$.
Moreover, because $\rho_1$ and $\rho_2$ are accepting, it must hold that $q_f^1 = p = q_f^2$, reaching a contradiction.

Now, consider a word $w = a_1\ldots a_n$ and the only accepting run
\[
\rho = (q_0,\emptyset) \trans[s_1]{a_1/\emptyset} (q_1,T_1) \trans[s_2]{a_2/\emptyset} \cdots \trans[s_n]{a_n/\emptyset}(q_n,T_n)
\]
where each $s_i$ is the cost of the $i$-th transition.
It is not hard to see that, for every $i \in [n]$, $s_i = \sem{\alpha'}(w,\sigma[x \rightarrow i])$ and, therefore,
\[
\sem{\alpha}(w,\sigma) \ = \ \sum_{i = 1}^n s_i \ = \ \costt{\cT}(w,\sigma)
\]

Now that we have a CT $\cT_\alpha = (Q_\alpha,\Delta_\alpha,\kappa_\alpha,I_\alpha,F_\alpha)$ and a DFA $\cA_\varphi = (Q_\varphi,\delta_\varphi,\kappa_\varphi,q_0^\varphi,F_\varphi)$ (over words in $\Sigma_V$), we build our final CT $\cT = (Q,\Delta,\kappa,I,F)$ by doing the cross-product of both as follows:
\begin{itemize}
	\item $Q = Q_\alpha \times Q_\varphi$
	\item $\Delta = \{((p_1,p_2),a,V',(q_1,q_2)) \mid (p_1,a,V',q_1) \in \Delta_\alpha \land (p_2,(a,V'),q_2) \in \delta_\varphi\}$
	\item $\kappa(((p_1,p_2),a,V',(q_1,q_2))) = \kappa_\alpha((p_1,a,V',q_1))$
	\item $I(q_1,q^\varphi_0) = I_\alpha(q_1)$ for all $q_1 \in \dom(I_\alpha)$
	\item $F(q_1,q_2) = F_\alpha(q_1)$ for all $(q_1,q_2) \in \dom(F_\alpha) \times F_\varphi$
\end{itemize}
From the above, one can check that $\sem{\varphi} = \sem{\cT}$ and $\sem{\alpha}(w,\sigma) = \costt{\cT}(w,\sigma)$ for every word $w$ and assignment $\sigma \in \sem{\varphi}(w)$.

\section{Complexity of Heap of Words operations}
\label{apx:how}
We recall $\hmeld$ and $\hincreaseby$ are implemented using their
incremental Brodal queue equivalent and thus inherit the constant time
complexity of those. We now recall the implementation of the other
operations on HoW:

\begin{algorithm}[h]
	\caption{\hnameshort{}'s implementation of $\hadd$, $\hextendby$, $\hfindmin$ and $\hdeletemin$.} 
	\begin{varwidth}[t]{0.55\textwidth}
		\begin{algorithmic}[1]
			\Procedure{$\hadd$}{$\hq{\bQ},\qpair{a}{g}$}
			\State \Return $\hq{\madd(\bQ,\qpair{(a,\hq{\emptyset})}{g})}$
			\EndProcedure
			
			\smallskip

			\Procedure{$\hextendby$}{$\hq{\bQ},a$}
			\If {$\misempty(\bQ)$}
			\State \Return $\hq{\emptyset}$
			\EndIf
			\State \Return $\hq{\madd(\emptyset,\qpair{(a,\hq{\bQ})}{\mminPriority(\bQ)})}$
			\EndProcedure
			\smallskip
			
			\Procedure{$\hfindmin$}{$\hq{\bQ}$}
			\State  $(a,\hq{\bQ'}) \gets \mfindmin(\bQ)$
			\If{$\misempty(\bQ')$}
			\State \Return $a$
			\EndIf
			\State \Return $\hfindmin(\hq{\bQ'}) \cdot a$
			\EndProcedure
			\algstore{myalg}
		\end{algorithmic}
	\end{varwidth}
	\quad
	\begin{varwidth}[t]{0.45\textwidth}
		\begin{algorithmic}[1]
			\algrestore{myalg}
			\Procedure{$\hdeletemin$}{$\hq{\bQ}$}
			\If {$\misempty(\bQ)$}
			\State \Return $\hq{\emptyset}$
			\EndIf
			\State  $(a,\hq{\bR}) \gets \mfindmin(\bQ)$
			\State  $\bQ' \gets \mdeletemin(\bQ)$
			\State $\hq{\bR'}  \gets \hdeletemin(\hq{\bR})$
			\If{$\iisempty(\bR')$}
			\State \Return $\hq{\bQ'}$
			\EndIf
			\State $\delta \gets \mminPriority(\bR')\gop (\mminPriority(\bR))^{\minus 1}$
			\State $g \gets \mminPriority(\bQ)\gop\delta$
			\State \Return $\hq{\madd(\bQ',\qpair{(a,\hq{\bR'})}{g})}$
			\EndProcedure
		\end{algorithmic}
	\end{varwidth}
\end{algorithm}

As we see the functions $\hadd$ and $\hextendby$ makes a constant
number of calls to the constant time functions $\madd$, $\misempty$
and $\mminPriority$ and thus they are constant time.

The function $\hfindmin$ is recursive but it will make one recursive
call for each letter in the output and at each recursive step it will
make one call to $\mfindmin$ and to $\misempty$, therefore the overall
complexity of $\hfindmin$ is linear in the returned word. Notice that
we use the $\cdot$ operator (to denote concatenation) on strings and
we suppose that it takes constant time. For this, we can encode
strings as lists of individual letters in reversed order and thus
appending a letter at the end of the word corresponds to appending at
the beginning of a list and this takes constant time.

Let us now dive into the complexity of $\hdeletemin$. We claimed that
the complexity of $\hdeletemin(h)$ was $\cO(|w|\times log(n))$ where
$w=\hfindmin(h)$ and $n$ is the number of operations that were used to
build $h$ without counting the $\hdeletemin$ (these $\hdeletemin$
should only happen at the enumeration phase, hence after all the
$\hadd$, $\hincreaseby$, $\hmeld$ and $\hextendby$).  The complexity
of $\hdeletemin$ is dominated by the calls to $\mdeletemin$, and there
are $|w|$ such calls. We thus need to prove that each of these
$\mdeletemin$ takes $\cO(log(n))$ time.

The \emph{branching factor} of a HoW $\hq{\bQ}$, noted
$\text{branch}(\hq{\bQ})$, is recursively defined as the maximum
between the number of elements in $\bQ$ and the branching factors of
HoW contained in $\bQ$. By definition, any $\mdeletemin$ operation
triggered by $\hdeletemin(h)$ will take a time at msot logarithmic in
$\text{branch}(h)$, thus it suffices to prove that
$\text{branch}(h)\leq n$

Let $\hstruct_0, \hstruct_1, \ldots, \hstruct_k$ be a sequence of
\hnameshortp{} such that $\hstruct_0 = \emptyset$ and each
$\hstruct_{i}$ is the result of one of the functions $\hadd$,
$\hincreaseby$, $\hextendby$, or $\hmeld$ applied over any of the
previous \hnameshortp{} $\hstruct_0, \ldots, \hstruct_{i-1}$. We will
prove that for all $i$ the queue $h_i$ contains less than $i$
elements.
Let us introduce $E_i$ as the set of pairs $(\hq{\bQ},a)$ for which
there exists at least one $g$ such that $\qpair{(a,\hq{\bQ})}{g}$
belongs to one of the queues $h_0, \ldots, h_{i-1}$. When $h_i$ is
built with $\hadd$ or $\hextendby$ we have that $|E_{i+1}\setminus
E_i|\leq 1$, when $h_i$ is built with $\hmeld$ or $\hincreaseby$ we
have $E_{i+1}=E_i$, all in all, we get $|E_i| \leq i$. Now, clearly
the queue $h_i$ contains less than $|E_i|$ elements and it is easy to
see that $\text{branch}(h_n)$ is bounded by $max_{i\leq
n}(|\hstruct_i|)$ where $|\hstruct_i|$ is the number of elements in
the queue of $\hstruct_n$ (queues that are not top queues are created
by an $\hextendby$ operation and never modified). Thus, we do have the
expected $\text{branch}(h_n)\leq n$ for an $h_n$ built without using
any $\hdeletemin$.

When applying $\hdeletemin$ the branching factor can only decrease:
$\hdeletemin(h)$ might add an edge to some of the queues in $h$ but
only after removing one in those queues. Therefore if $h$ is obtained
by using $n$ operations $\hadd$, $\hincreaseby$, $\hmeld$ and
$\hextendby$ followed by $k$ operations $\hdeletemin$, its branching
factor is bounded by $n$. And that gives us the complexities
requiered in Section~\ref{sec:renumalgo}.

\section{Building incremental Brodal queues} 

\newcommand{\bP}{P}

Here we continue the work of Section~\ref{sec:queues} and present the complete incremental Brodal queue.
In the following, we use a similar approach as the one used in~\cite{brodal1996optimal}: we start with the incremental binomial heap, which supports most operations in $\cO(\log n)$; then, we explain each modification made to this structure until we have built the final structure.

\subsection{Skew incremental binomial heap}
We now explain how the time of $\madd$ can be reduced to $\cO(1)$, while maintaining the asymptotic time of the other operations.
The technique is borrowed from~\cite{brodal1996optimal} and modified to handle $\mincreaseby$ efficiently.

The motivation comes from the skew binary numbers~\cite{myers1983applicative}, in particular the \emph{canonical} skew binary numbers, a variation of binary numbers in which all digits are $0$ or $1$ with the possible exception of the lowest order non-zero digit, which might be $2$.
A skew binary number $\alpha = \alpha_n \alpha_{n-1} \ldots \alpha_1$ denotes the integer value $\sum^n_{i=1}a_i(2^i-1)$.
This number representation avoids the carry cascading when adding $1$ to a number.
For example, the number $43$ is represented by the skew binary number $10112$, and adding $1$ to it results in $10120$, which is done with a single carry operation.

A \emph{skew binomial tree}, or skew tree, is a tree with the following definition:
\begin{itemize}
	\item a skew tree of rank $0$ is a leaf.
	\item a skew tree of rank $r+1$ is formed in one of three ways:
	\begin{itemize}
		\item a \emph{simple link}, making a skew tree of rank $r$ the leftmost child of another skew tree of rank $r$;
		\item a \emph{type A link}, making two skew trees of rank $r$ the children of a skew tree of rank $0$; or
		\item a \emph{type B link}, making a skew tree of rank $0$ and a skew tree of rank $r$ the leftmost children of another skew tree of rank $r$.
	\end{itemize} 
\end{itemize}
Note that, unlike binomial trees, now a skew binomial tree of rank $r$ has a less rigid structure and, in particular, the number of contained nodes $n$ is not fixed.
However, is not hard to see that it is bounded by $2^r \leq n \leq 2^{r+1}-1$.
Moreover, given a tree $\tT$ with root $v_\tT$ and $\trank(\tT) = r$, the number of children of $v_\tT$ is proportional to $r$, and therefore $\chd(v_\tT)$ is $\cO(\log |V_{v_\tT}|)$.

A \emph{skew incremental binomial heap} (skew heap for short) is defined as a tuple $\h = (\tdom, \fch, \nsb, \troot, \tdelta, \tval, \tdeltainit)$, where all components are the same as for incremental binomial heaps, except that if $\troots = v_1, \ldots, v_n$ are its roots, then each $\M_{v_i}$ is an incremental skew binomial tree with $\trank(\M_{v_i}) < \trank(\M_{v_{i+1}})$ for each $2 < i < n$, and $\trank(\M_{v_1}) \leq \trank(\M_{v_2})$.
Namely, we allow the two smallest trees to have the same rank.

Consider a skew heap $\h = (\tdom, \fch, \nsb, \troot, \tdelta, \tval, \tdeltainit)$ with $\troots = v_1, \ldots, v_n$.
Functions $\misempty$, $\mincreaseby$, $\mfindmin$, $\mminPriority$ and $\mmeld$ remain the same as for regular heaps.

For $\madd(\h,e,g)$, we create a new skew heap $\h'$ with $\h \subseteq \h'$ as follows. 
Let $v$ be a fresh node with $\tdeltai{\h'}(v) := g \gop \tdelta^{-1}$ and $\tvali{\h'}(v) := e$.
If all trees $\M_{v_i}$ have different rank, then we add $v$ at the beginning of the roots, namely $v$ is made the root of $\h'$ and $\nsb_{\h'}(v) = \troot$.
If there are two trees $\M_{v_1}$, $\M_{v_2}$ with the same rank, we \emph{link} $v$ with $v_1$ and $v_2$ to form a tree of rank $r+1$ as follows.
If $\tdelta(v) < \tdelta(v_1)$ and $\tdelta(v) < \tdelta(v_2)$, then, we make copies $v_1',v_2'$ of $v_1$ and $v_2$ with their priorities decreased by $\tdelta_{\h'}(v)$,  i.e. $\tdelta_{\h'}(v_i) = \tdelta_{\h}(v_i)\gop \tdelta_{\h'}(v)^{-1}$, make them the children of $v$ and make $v$ the root of $\h'$ with $\nsb_{\h'}(v) = \nsb_{\h}(v_2)$.
Otherwise, w.l.o.g. consider that $\tdelta(v_1) < \tdelta(v_2)$.
Then, we make copies $v_1',v_2'$ of $v_1$ and $v_2$ and make $v_2'$ and $v$ the leftmost children of $v_1'$, but decreasing their priorities by $\tdelta(v_1')$ in the same way.
Note that the first case represents an A link and the latter a B link, and that the new tree is a valid skew tree of rank $r+1$ that stores the same elements as $v_1$ and $v_2$, plus the new element $v$.
Moreover, since we allow two repeated ranks and we know that there is at most one other tree of rank $r+1$, adding it does not require a chain of links and thus takes time $\cO(1)$.

The case of $\mdeletemin$ follows the same principle: we make a skew heap $\h_1$ to store copies of the roots except for the minimal one, call it $v_i$, make a skew heap $\h_2$ to store the children of $v_i$, and build the result with $\mmeld(\h_1,\h_2)$.
However, building $\h_2$ requires a little more work.
Since $\M(v_i)$ is a skew heap, among their children could appear up to $\log n$ skew heaps of rank $0$.
Then, we build $\h_2$ by taking the children with rank higher than $0$, and then add the remaining ones using $\madd$, taking time $\cO(\log n)$.
Finally, we return $\mmeld(\h_1,\h_2)$, taking overall time $\cO(\log n)$.

\subsection{Incremental Brodal queue}
Clearly, this structure alone is not enough for what we need, since most operations still take time $\cO(\log n)$.
To address this, we base one more time on the techniques of~\cite{brodal1996optimal}, this time on one called ``bootstrapping''.

An \emph{incremental Brodal queue} (\emph{queue} for short) is either the empty queue $\bQ_\emptyset$ or a pair $\bQ = ((e,p),\h)$, where $e$ is a stored element, $p$ is its priority and $\h$ is a skew heap that stores incremental Brodal queues.
Given a queue $\bQ$, we denote its components with the subscript $\bQ$, e.g., $e_{\bQ}$, $p_{\bQ}$, $\h_{\bQ}$.
Consider some $(\bQ,p)$ stored in $\h$.
Note that, whenever $\h$ is applied an $\mincreaseby$ operation, $p$ is increased but the priorities inside $\bQ$ are not; instead, we increase them only when $(\bQ,p)$ is retrieved from $\h$.
We use this setting because the $\mincreaseby$ operation is able to increase priorities stored in $\h$ efficiently, but increasing all priorities inside $\bQ$ might result in an unbounded recursion.
Priority $p$ is then going to store the real updated value of $p_{\bQ}$ considering how much the priorities of $\bQ$ have been increased by calling $\mincreaseby$ over $\h$.
Note that $p$ can be seen as having the form $p_\bQ \gop d$ for some $d \in \bbG$ that represents such increase and, thus, $d = p_\bQ^{-1} \gop p$.
Consequently, each priority in $\bQ$ actually represents a priority $p \gop d$.
The set of elements stored in a skew heap $\h$ and a queue $\bQ$ are defined accordingly by the following two-level recursive definition:
\begin{align*}
\sem{\h} & = \bigcup_{(\bQ,p) \in \set{\h}} \{\sem{\bQ}_{\gop (p_{\bQ}^{-1} \gop p)}\}\\
\sem{\bQ} & = \{(e_{\bQ},p_{\bQ})\} \cup \sem{\h}
\end{align*}
where $\sem{\bQ}_{\gop d} = \{(e,p\gop d) \mid (e,p) \in \sem{\bQ}\}$ represents that all priorities of $\sem{\bQ}$ are increased by $d$.
This recursive definition just says that the $\sem{\bQ}$ consists of all elements stored in $\bQ$, including the ones at other queues inside $\bQ$.
Moreover, $\bQ$ is kept so that $(e_\bQ,p_\bQ)$ is always the element with minimum priority of $\sem{\bQ}$.

Consider a queue $\bQ = ((e_{\bQ},p_{\bQ}),\h_{\bQ})$, and let $\bP = ((e_{\bP},p_{\bP}),\h_{\bP})$ be the resulting queue after applying each operation over $\bQ$.
Then, $\bQ$ implements the same operations as a heap:
\begin{itemize}
	\item $\misempty(\bQ)$.
	It only checks if $\bQ$ is equal to the empty queue $\bQ_\emptyset$ and returns accordingly.
	
	\item $\madd(\bQ,\qpair{e'}{p'})$.
	First, if $\bQ$ is empty, we initialize a new queue $\bP$ with $(e_{\bP},p_{\bP}) = (e', p')$ and $\h_{\bP} = \h_\emptyset$, where $\h_\emptyset$ is the empty heap.
	Otherwise, if $p' < p_{\bQ}$, then $e'$ is the new minimum, so we set $(e_{\bP},p_{\bP}) = (e',p')$, otherwise we set it to $(e_{\bQ},p_{\bQ})$.
	In either case, we set $\h_{\bP} = \h_{\bQ}$.
	W.l.o.g., let $(e',p')$ be the one that is not the minimum.
	In order to add $(e',p')$ to $\h_{\bP}$, we create a new queue $\bQ'=((e',p'),\h_\emptyset)$, and then add $\bQ'$ to $\h_{\bP}$ with $\madd(\h_{\bP},\qpair{\bQ'}{p'})$.
	
	\item $\mfindmin(\bQ)$ and $\mminPriority(\bQ)$.
	We just return $e_{\bQ}$ or $p_{\bQ}$, respectively, and keep $\bP = \bQ$.
	
	\item $\mdeletemin(\bQ)$.
	It is easy to find $(e_{\bQ},p_{\bQ})$, but to delete it we need to replace it with the next minimum from $\h_{\bQ}$ afterwards.
	We find it by running $\bR = \mfindmin(\h_{\bQ})$, $u = \mminPriority(\h_{\bQ})$, and then delete it with $\h_{\bP} = \mdeletemin(\h_{\bQ})$.
	Consider that the queue $\bR$ has the form $((e_{\bR},p_{\bR}),\h_{\bR})$ and recall that $u$ stores the updated value of $p_{\bR}$.
	Then, the minimum element is $(e_{\bR},u)$, so we set that as our new $(e_{\bP},p_{\bP})$.
	Notice that we removed queue $\bR$, meaning that we not only removed the minimum element, but also all elements stored in $\h_{\bR}$.
	In order to add them again, we first increase the priorities in $\h_{\bR}$ by $(p_{\bR}^{-1} \gop u)$ and then we meld it with $\h_{\bP}$, by doing $\h_{\bP} = \mmeld(\mincreaseby(\h_{\bR},p_{\bR}^{-1} \gop u),\h_{\bP})$.
	
	\item $\mmeld(\bQ,\bR)$.
	Let queue $\bR$ have the form $((e_{\bR},p_{\bR}),\h_{\bR})$.
	First, we select between $(e_{\bQ},p_{\bQ})$ and $(e_{\bR},p_{\bR})$ the one with the lowest priority; assume w.l.o.g. that it is $(e_{\bQ},p_{\bQ})$.
	Then, we set $(e_{\bP},p_{\bP}) = (e_{\bQ},p_{\bQ})$ and $\h_{\bP} = \madd(\h_{\bQ},\qpair{\bR}{p_{\bR}})$.
	
	\item $\mincreaseby(\bQ,g)$.
	We set $e_{\bP} = e_{\bQ}$, $p_{\bP} = p_{\bQ} \gop g$ and $\h_{\bP} = \mincreaseby(\h_{\bQ},g)$, that is, just increase $p_{\bQ}$ by $g$ and apply $\mincreaseby(\h,g)$.
	Recall that the latter increases only the priorities stored directly in $\h_{\bQ}$, namely each $(\bQ',p')$ in $\h_{\bQ}$ turns into $(\bQ',p'\gop g)$ in $\h_{\bP}$.
	The priorities of elements inside $\bQ'$ remain unchanged, and are updated later when $\bQ'$ is retrieved from $\h_{\bP}$.
	
\end{itemize}
Again, the structure is fully persistent because none of the operations modify $\bQ$.
It is not hard to see that now operations $\madd$, $\mfindmin$, $\mmeld$ and $\mincreaseby$ take time $\cO(1)$, and $\mdeletemin$ takes time $\cO(\log n)$.

\end{document}